\documentclass{emulateapj}
\shortauthors{Gabel et al.}
\shorttitle{Variability and Modeling the UV Absorption in NGC 3783}

\begin{document}
\title{The Ionized Gas and Nuclear Environment in NGC 3783\\
V. Variability and Modeling of the Intrinsic Ultraviolet Absorption\altaffilmark{1}}
\author{Jack R. Gabel\altaffilmark{2}, Steven B. Kraemer\altaffilmark{3},
D. Michael Crenshaw\altaffilmark{4},  Ian M. George\altaffilmark{5,6},
W. N. Brandt\altaffilmark{7},
Frederick W. Hamann\altaffilmark{8}, Mary Elizabeth Kaiser\altaffilmark{9},
Shai Kaspi\altaffilmark{10,11}, 
Gerard A. Kriss\altaffilmark{9,12}, Smita Mathur\altaffilmark{13}, 
Kirpal Nandra\altaffilmark{14}, Hagai Netzer\altaffilmark{11},
Bradley M. Peterson\altaffilmark{13}, Joseph C. Shields\altaffilmark{15},
T. J. Turner\altaffilmark{5,6}, \& Wei Zheng\altaffilmark{9}}
\altaffiltext{1}{Based on observations made with the NASA/ESA {\it Hubble
Space Telescope} obtained at the Space Telescope Science Institute, which
is operated by the Association of Universities for Research in Astronomy, Inc.
under NASA contract NAS~5-26555, and with the NASA-CNES-CSA {\it Far Ultraviolet Spectroscopic
Explorer}, which is operated for NASA by the Johns Hopkins University under NASA contract
NAS5-32985.}
\altaffiltext{2}{Center for Astrophysics and Space Astronomy, University of Colorado, 389 UCB,
Boulder CO 80309-0389; jgabel@colorado.edu}
\altaffiltext{3}{The Catholic University of America/IACS, NASA/Goddard Space Flight Center,
Laboratory for Astronomy and Solar Physics, Code 681, Greenbelt, MD 20771.}
\altaffiltext{4}{Department of Physics and Astronomy, Georgia State University,
Atlanta, GA 30303.}
\altaffiltext{5}{Laboratory for High Energy Astrophysics, NASA/Goddard Space Flight Center,
Code 662, Greenbelt, MD 20771.}
\altaffiltext{6}{Joint Center for Astrophysics, Physics Department, University of Maryland,
Baltimore County, 1000 Hilltop Circle, Baltimore, MD 21250.}
\altaffiltext{7}{Department of Astronomy and Astrophysics, 525 Davey Laboratory, The
Pennsylvania State University, University Park, PA 16802.}
\altaffiltext{8}{Department of Astronomy, University of Florida, 211 Bryant Space Science Center,
Gainesville, FL, 32611-2055.}
\altaffiltext{9}{Center for Astrophysical Sciences, Department of Physics and Astronomy,
The Johns Hopkins University, Baltimore, MD 21218-2686.}
\altaffiltext{10}{Physics Department, Technion, Haifa 3200, Israel.}
\altaffiltext{11}{School of Physics and Astronomy, Raymond and Beverly Sackler Faculty of
Exact Sciences, Tel-Aviv University, Tel-Aviv 69978, Israel.}
\altaffiltext{12}{Space Telescope Science Institute, 3700 San Martin Drive, Baltimore, MD 21218.}
\altaffiltext{13}{Department of Astronomy, Ohio State University, 140 West 18th Avenue,
Columbus, OH 43210-1173.}
\altaffiltext{14}{Astrophysics Group, Imperial College London, Blackett Laboratory, Prince Consort Road, 
London SW7 2AW.}
\altaffiltext{15}{Department of Physics and Astronomy, Clippinger Research Labs 251B, Ohio
University, Athens,  OH 45701-2979.}

\begin{abstract}
We present results on the location, physical conditions, and geometry
of the outflow in the Seyfert 1 galaxy NGC 3783 from a study of the variable 
intrinsic UV absorption.
Based on analysis of 18 observations with the Space Telescope Imaging Spectrograph aboard
the {\it Hubble Space Telescope} and 6 observations with the 
{\it Far Ultraviolet Spectroscopic Explorer} obtained between 2000 February and 2002 January, 
we obtain the following results:
1) The lowest-ionization species detected in each of the three strong kinematic components
(components 1 -- 3 at radial velocities $-$1350, $-$550, and $-$725 km~s$^{-1}$, respectively)
varied, with equivalent widths inversely correlated with the continuum
flux.  This indicates the ionization
structure in the absorbers responded to changes in the photoionizing flux, with
variations occurring over the weekly timescales sampled by our observations.
2) A multi-component model of the line-of-sight absorption covering factors, 
which includes an unocculted narrow emission-line region (NLR) and 
separate covering factors derived for the broad line region 
and continuum emission sources, predicts saturation in several lines, consistent
with the lack of observed variability in these lines.
Differences in covering factors and kinematic structure imply component~1 is
comprised of two physically distinct regions (1a and 1b).
3)  We obtain column densities for the individual metastable levels from
the resolved \ion{C}{3}*~$\lambda$1175 absorption complex in component~1a.
Based on our computed metastable level populations, the electron density of
this absorber is $\sim$ 3$\times$10$^4$~cm$^{-3}$.
Combined with photoionization modeling results, this places component~1a at
$\sim$~25~pc from the central source.
5)  Using time-dependent calculations, we are able to reproduce the
detailed variability observed in component~1 and derive upper limits on the
distances for components 2 and 3 of $\leq$ 25 and $\leq$ 50 pc, respectively.
6) The ionization parameters derived for the higher ionization UV absorbers (components 1b,
2, and 3 with log($U$)$\approx-$0.5) are consistent with the modeling results
for the lowest-ionization X-ray component, but with smaller total
column density. The high-ionization UV components are found to have similar pressures as
the three X-ray ionization components.
These results are consistent with an inhomogeneous wind model for the outflow
in NGC~3783, with denser, colder, lower-ionization regions embedded in more highly-ionized gas. 
7)  Based on the predicted emission-line luminosities, global covering factor constraints,
and distances derived for the UV absorbers, they may be identified with emission-line gas
observed in the inner NLR of AGNs.
We explore constraints for dynamical models of AGN outflows implied by
these results.
\end{abstract}

\section{Introduction}

    Mass outflow, seen as blueshifted absorption in UV and X-ray spectra,
is an important component of active galactic nuclei \citep[see the recent
review in][]{cren03}.
This ``intrinsic absorption" is ubiquitous in nearby AGNs, appearing
in over half of Seyfert 1 galaxies having high-quality UV spectra obtained
with the {\it Hubble Space Telescope} ({\it HST}) \citep{cren99}
and the {\it Far Ultraviolet Spectroscopic Explorer} ({\it FUSE}) \citep{kris02}.
Spectra from the {\it Advanced Satellite for Cosmology and Astrophysics (ASCA)}
identified X-ray ``warm absorbers", modeled as absorption edges, in a similar percentage of objects
\citep{reyn97,geor98}, and many studies have explored the connection
between the absorption observed in the X-ray and UV bandpasses \citep[e.g.][]{math94}.
Large total ejected masses have been inferred for these
outflows, exceeding the accretion rate of the central black hole in some cases,
indicating mass outflow may play an important role in the overall energetics in
AGNs \citep[e.g.][]{math95,reyn97}.

   Variability in the intrinsic UV absorption in Seyfert galaxies is common.
All objects with high-resolution UV spectra obtained at multiple epochs exhibit
substantial variations in their absorption strengths \citep[][and references therein]{cren03}.
Absorption variability could result from: (a) a response to changes
in the ionizing AGN flux such that the total column density of the absorber
remains constant but the ionization structure changes, 
(b) regions of condensation/evaporation in our line-of-sight to the AGN emission sources 
due to thermal perturbations, or
(c) bulk motion of the absorber transverse to our line-of-sight.  In the latter case, the observed equivalent widths could vary due to either
a change in the line-of-sight covering factor of the background emission,
or a change in the total column density of gas seen due to a shifting
of different regions of the outflow across our sightline.
In each case, the measured variability characteristics provide important constraints on the
dynamics, geometry, and/or physical state of the AGN absorbers.
For motion across the background AGN emission, the variability timescales can
constrain the transverse component of the kinematics of the absorber and the absorption-emission
geometry.  For ionization changes, the observed absorption variability can be used to constrain the
gas number density and, combined with photoionization models, determine the distance of
the absorber from the central source.
These parameters are needed to determine the mechanism driving the mass outflow and the
source of the absorption gas, and assess its overall role in the energetics of the AGN.

  The bright Seyfert 1 galaxy NGC 3783 has a rich UV and X-ray absorption spectrum.
High-resolution observations with the Goddard High Resolution Spectrograph (GHRS) 
and the Space Telescope Imaging Spectrograph (STIS) aboard {\it HST} showed the 
UV absorption is highly variable.
Three distinct kinematic components of absorption appeared independently over
yearly timescales:
components 1 -- 3 having radial velocities $v_r \approx -$1350, $-$550, and $-$725~km~s$^{-1}$
and widths $FWHM \approx$ 190, 170, and 280~km~s$^{-1}$ 
\citep[][hereafter KC01]{krae01}.\footnote{Tentative detection of a weak, fourth component at 
$v_r \approx -$1050~km~s$^{-1}$ was described in \citet{gabe03a}; 
we do not treat this component in the following analysis.}
These long-term changes were found to be inconsistent with ionization changes,
and thus interpreted as a signature of transverse motion of the absorbers by KC01.
High-resolution X-ray observations with the {\it Chandra X-ray Observatory (CXO)} 
revealed a spectrum with numerous absorption lines from a large range in ionization 
states \citep{kasp01}.

    We have undertaken an intensive, multiwavelength campaign on NGC 3783 with
{\it HST}/STIS, {\it FUSE}, and {\it CXO} to monitor the absorption properties.
Earlier papers in this series have presented studies of the mean X-ray \citep[][ Paper I]{kasp02} 
and UV \citep[][ Paper II]{gabe03a} absorption spectra, analysis of a decrease
in radial velocity detected in UV component~1 \citep[][ Paper III]{gabe03b},
and variability and detailed modeling of the X-ray absorption \citep[][ Paper IV]{netz03}.
In this paper, we present a study of the variability and physical conditions in the UV absorption
based on analysis of the {\it FUSE} and STIS spectra.
In \S 2, we review the observations and present the UV continuum light curve;
in \S 3, we present measurements of the absorption parameters and variability.
We analyze metastable \ion{C}{3}*~$\lambda$1175 absorption detected in component~1 in \S 4 to
derive the number density in this absorber.
In \S 5, we present detailed modeling of the UV absorption, making use of observed
variability in the spectrum, and explore time-dependent solutions 
in response to the continuum variations.
We explore global models of the outflow in NGC 3783 in \S 6, based on the combined results
from the UV and X-ray analysis.

\section{Observations and the UV Continuum Light Curve}

\subsection{{\it FUSE} and {\it HST}/STIS Echelle Spectra}

  We present results from a total of 18 medium-resolution STIS echelle spectra
(S1 -- S18) and 6 {\it FUSE} spectra (F1 -- F6) of the nucleus of NGC 3783, obtained
between 2000 February and 2002 January.
A detailed description of the observations and data reduction is given in Paper II;
here we present a brief overview.

    Each STIS observation was obtained using the 0$\farcs$2~$\times$~0$\farcs$2
aperture and the E140M grating, which spans 1150--1730 \AA, and consisted of two
{\it HST} orbits for a total exposure time of $\sim$ 4.9 ks (except S1 which was
5.4 ks).  The STIS spectra were processed with IDL software developed at NASA's Goddard
Space Flight Center for the Instrument Definition Team, which includes a procedure
to remove background light from each order using a scattered light model
devised by \citet{lind99}.  Our measurements of the residual fluxes in the cores of saturated
interstellar Galactic lines show the scattered light was accurately removed (see Paper~II).
The extracted STIS spectra are sampled in $\sim$~0.017~\AA~bins, thereby preserving
the full resolution of the STIS/E140M grating ($FWHM~\approx$~7~km~s$^{-1}$).

    The {\it FUSE} spectra, covering 905 -- 1187 \AA, were obtained through the
30$\arcsec$~$\times$~30$\arcsec$ aperture.
Each spectrum was processed with the standard calibration pipeline, CALFUSE.
For each observation, the eight individual spectra obtained with
{\it FUSE} from the combination of four mirror/grating channels and two
detectors, were coadded for all exposures.  
We corrected for nonlinear shifts in wavelength scale between the spectra
from different detector segments by cross-correlating over small bandpasses.
The absolute wavelength scale was determined by matching the velocities of Galactic
lines in the {\it FUSE} spectrum with those in the STIS bandpass.
Mean residual fluxes measured in the cores of
saturated Galactic lines are consistent with zero within the noise (i.e., standard
deviation of the fluxes) in the troughs of these lines, indicating accurate background removal
for all observations except F5.  For this observation, we fit the remaining residual background flux in
strong interstellar lines to match the other epochs,  and subtracted the fit.
The spectra were resampled into $\sim$ 0.02 -- 0.03 \AA~bins to increase the signal-to-noise
ratio (S/N) while preserving the full resolution of {\it FUSE}, which is nominally
$FWHM~\approx$~20~km~s$^{-1}$.

\subsection{The UV Continuum Light Curve}

  To put the continuum variations of our observations in perspective,
Figure 1a shows the continuum flux light curve at 1470~\AA~for
all UV spectra of NGC 3783 obtained over the last 22 years.
These observations were obtained with the
{\it International Ultraviolet Explorer} ({\it IUE}) and the Faint Object
Spectrograph (FOS), GHRS, and STIS (in low dispersion) on {\it HST}, which are listed in Table~1.
We obtained the most recently processed version of each spectrum from the Multimission
Archive at the Space Telescope Science Institute.  
We measured the continuum fluxes in each spectrum by averaging the elements in a
30~\AA~bin centered at 1470~\AA~in the observed frame, which is free of contamination from
line emission.
The 1$\sigma$ flux uncertainties were determined from the standard deviations.
For the {\it IUE} spectra, this technique is known to overestimate the errors \citep{clav91}; 
therefore, following the procedure described in \citet{krae02}, 
we scaled these uncertainties by a factor
of 0.5 to ensure that observations taken on the same day agreed to within the
errors on average. 
Due to the small wavelength coverage of the GHRS spectra, 
no continuum regions free of broad line emission were observed.
To estimate the GHRS continuum fluxes, we used separate
fits to the continua and broad emission lines from the STIS spectra and tested
different linear combinations of these fits until we obtained an accurate
match to the observed GHRS profiles.
Figure~1a shows the STIS echelle observations (JD~$\approx$~2,452,000) sampled the UV
continuum in a range of moderately-low to moderately-high flux states compared
to the long-term light curve.

    In Figure~1b, the light curve for the monitoring campaign observations is
shown in more detail.
We have included estimated fluxes from the four {\it FUSE} observations that
were not simultaneous with STIS observations (F1, F3, F5, F6) by extrapolating
from the far-UV bandpass to 1470~\AA.  The first four observations (S1 -- S4)
were obtained at intervals of several months, the intensive monitoring phase (S5 -- S17)
sampled the continuum at 3 -- 8 day intervals, and the final observation (S18) 
was obtained about 9 months later.
The extrema in continuum flux in the STIS observations occurred during the long-term 
sampling;
S2 and S3 observed the continuum in the highest flux state and the final observation found it
in the lowest state, with a peak amplitude variation of a factor of $\sim$2.5.
We identify two general phases during the intensive monitoring: 
the first four observations (S5 -- S8) sampled the continuum in
a relatively low-state, after which it increased by up to a factor 
of 1.7 (S10 and S12) over a period of about 2 weeks.
Three {\it FUSE} observations (F2 -- F4) were obtained during the low-state
STIS observations, while observation F5 observed the continuum in the highest 
state (four days after S12).
In the subsequent variability analysis, we compare observations
between mean low and high states derived by averaging representative spectra in each
state.

\section{Measurements of Absorption Parameters}

  A key issue in the analysis of AGN outflows is extracting accurate
ionic column densities from the observed absorption lines, since they
provide the basis for determining the physical conditions in the absorbers 
(i.e., the ionization state, total gas column, number density).
The UV absorbers typically only partially occult the background
AGN emission, as determined by the relative strengths of the individual
members of absorption doublets \citep{wamp93,hama97,barl97}, and the
effects of covering factor ($C$) and optical depth ($\tau$) must be deconvolved to
determine the column densities.  In some cases, the $C - \tau$ solution for observed
features can be complicated due to blending of physically distinct absorption components
or complex coverage of the background emission.  Thus, it is useful to first explore absorption
variability by comparing equivalent widths.

\subsection{Variability in Absorption Equivalent Widths}

   Figures 2a -- c show the equivalent widths for key lines in the STIS spectra of
components 1 -- 3, respectively, plotted as a function of the UV continuum flux.
The error bars represent our estimated measurement uncertainties, which are a
combination of uncertainties due to spectral noise and fitting the intrinsic
(i.e., unabsorbed) fluxes, via propagation of errors.
We estimated uncertainties in the intrinsic fluxes by testing different
empirical fits over the absorption features and selecting the range of what we deemed
to be reasonable line profile shapes. 
We note that these derived uncertainties may in some cases be too conservative,
as indicated by comparing the scatter in measured values with the error bars
plotted in Figure 2 (e.g., \ion{N}{5} in components 1 and 3). 
This may be due to systematic trends in how the intrinsic flux was fit in each
observation and/or overestimates of the uncertainties associated with the fitting.

    In each kinematic component, the lowest-ionization species with detectable absorption
other than \ion{H}{1} (\ion{Si}{4} in component~1;
\ion{C}{4} and \ion{N}{5} in component~2; \ion{C}{4} in component~3) show clear variations that are
inversely proportional to the continuum flux.  This is just as expected for unsaturated
lines from relatively low-ionization species in a photoionized gas, as their column densities 
decrease in higher flux states due to the increased ionization.  This gives direct
evidence that the ionization structure in these absorbers is dominated by
photoionization from the central source. It also indicates these lines are not
highly saturated, at least in the epochs with weaker observed absorption.
To test the dependence of absorption variability on continuum flux more rigorously,
we did a linear regression fit for each line.
Figure~2 shows the results of the fits and gives the ratio of the
computed slope to the 1$\sigma$ uncertainty in the slope, $m/\Delta m$,
to characterize the correlation.
Fits to all of the lines listed above have a non-zero slope at greater than a 3$\sigma$ level.
Since the absorption strengths will not necessarily vary linearly with continuum flux,
we also tested the Spearman rank correlation.
All of the above lines show correlation with the flux at high significance, 
with probabilities of no correlation computed to be between 0.01 -- 0.3 \%  from this 
test (values are listed in Figure 2).

   Conversely, the absorption strengths of \ion{C}{4} in component~1 and \ion{N}{5} in
components 1 and 3 did not vary strongly during the monitoring.
This could indicate either these lines were saturated, with partial covering factor, 
or their ionic column densities were not varying.
In the latter case, the lack of variability could be due to either a 
relatively low electron density such that the ionic populations are unable to 
respond to continuum changes, or to the ionic structure of the gas, with these ions near
the peak ionization states of their parent elements.  These possibilities are explored below.
We note that since each component has some lines that exhibit no variability (see Figure 2 and \S 3.2),
the lines that do vary must result from changes in ionic column densities rather than 
covering factors.

\subsection{Effective Line-of-Sight Covering Factors and Column Densities}

    The doublet method for measuring the absorption parameters
provides a solution to a single $C$ and $\tau$ (at each resolution
element in an absorption profile).  However, since the background AGN emission
is comprised of multiple, physically distinct components with different sizes
and geometries (i.e., a featureless continuum source and multiple kinematic
components of line emission), additional constraints may be needed to determine
the required {\it effective} covering factors, which are weighted combinations
of covering factors of the different emission sources (Ganguly et al. 1999).
Implicit in the doublet solution is that all emission sources have the same covering factor.
In Paper II, we used the Lyman series lines to separate the continuum and emission
line covering factors for the \ion{H}{1} absorption.  Here, we use variability in
the background AGN emission as an additional constraint to explore the effect of
the NLR on the derived covering factors and obtain a consistent model to measure absorption
column densities for all lines.

\subsubsection{Isolating the Narrow-Line Region Emission Profile}

   We first compare spectra in different flux states to isolate distinct
emission-line components based on variability in the overall profiles.
Figure 3a shows the \ion{C}{4} line profiles in mean
high-state (S2 and S3) and low-state (S5 -- S8) spectra; these observations were selected
for comparison because they show the largest difference in emission-line flux.
The continuum flux has been subtracted from each spectrum and
the low-state profile has been scaled by a factor of 1.4 to match the flux
in the high-velocity wings of the high-state spectrum.
The profiles match very well at radial velocities $|v_r| > $1500~km~s$^{-1}$
but diverge at lower velocities, with increasing discrepancy toward line center.
This effect is consistent with the superposition of a varying broad component and a non-varying
narrower emission-line component, hereafter the BLR and NLR respectively.

    Assuming no change in the NLR flux between observations, 
we isolate its profile by solving the following expression for $F_{NLR}$
at each radial velocity:
\begin{equation}
F_{low}(v) \times f_{sc} - F_{high}(v) = F_{NLR}(v) \times (f_{sc} - 1),
\end{equation}
where $F_{low}$ and $F_{high}$ are the total observed emission-line fluxes in each state (i.e., the
NLR $+$ BLR fluxes) and $f_{sc}$ is the scale factor equating the low-state BLR flux with the 
high-state.  The resulting NLR profile is plotted as a dotted line in Figure 3a. 
This analysis assumes the
BLR scales by a uniform factor at all radial velocities between states.
Equation~1 is not valid in spectral regions affected by absorption.
We have fit the NLR profile using Gaussians for each of the \ion{C}{4} doublet lines constrained
to have their intrinsic 2:1 flux ratio and velocity separation, giving a width of
$\sigma =$~500~$\pm$120~km~s$^{-1}$ ($FWHM =$~1180~$\pm$280~km~s$^{-1}$) at a radial velocity 
of $+$50~$\pm$60~km~s$^{-1}$ with respect to the systemic velocity. 
The combined fit, which appears symmetrical due to the broad line 
width relative to the separation of the doublet members, is plotted over the residual 
NLR flux in Figure 3a.

   We applied the same analysis to derive the NLR profiles of other emission lines using equation 1.
Results for Ly$\alpha$, \ion{N}{5}, and \ion{Si}{4} ($f_{sc}=$1.4, 1.8, and
1.8, respectively) are shown in Figure 3b.  A scaled template of the Gaussian fit to the \ion{C}{4}
NLR profile is overlaid on each residual NLR profile, including both lines for the
\ion{N}{5} and \ion{Si}{4} doublets.  The discrepancy in the residual fluxes in the
blue wing of \ion{N}{5} ($v_r < -$2000~km~s$^{-1}$) and the red wing of Ly$\alpha$ 
($v_r >$ 3000~km~s$^{-1}$)
is due to the different BLR flux scale factors for the two lines and the overlap in their BLR emission.
The excess emission in the \ion{Si}{4} profile between $\sim$ 1000 -- 2500 km s$^{-1}$ is due to the O IV]
emission-line multiplet.
The Ly$\alpha$ NLR is not well matched by the \ion{C}{4} template, appearing
narrower in the unabsorbed red wing of the profile.
Figure 3 shows the derived NLR fluxes contribute substantially at the wavelengths
of many of the absorption features and thus its covering factor could affect the
measurement of column densities, which we explore below.

   There are some caveats regarding the derivation of the NLR profiles.  
First, the assumption that the BLR scale factor between the low and high states is independent
of radial velocity may introduce an error into the solution.
Intensive UV -- optical monitoring of continuum and BLR variability in the Seyfert 1 galaxy NGC 5548 has 
revealed its BLR variability is consistent with a constant virial product 
\citep{pete99,pete04}.
A consequence of this is that the integrated BLR profile becomes narrower in higher
flux states; if this is the case for the UV lines in NGC 3783, then a velocity-dependent
scale factor in equation~1 would be required to fully separate the NLR and BLR.
Additionally, the profiles derived from this analysis are broader than are typical for
UV NLR lines in Seyfert 1 galaxies.  For example, {\it HST} observations of 
NGC 5548 and NGC 4151 while in low flux states with little contamination from 
BLR emission revealed $FWHM \approx$ 500 and 300~km~s$^{-1}$, respectively, for the \ion{C}{4} NLR.
However, the width derived for NGC 3783 is comparable to the rather broad NLR features in 
the Seyfert 2 galaxy NGC 1068 \citep{diet98,krae98a}.
To assess the NLR fit, we compare their flux ratios relative to the measured 
[\ion{O}{3}]~$\lambda$5007 line with those from other AGNs.
From optical spectra of NGC 3783 obtained with a 5$\arcsec$$\times$15$\arcsec$ 
aperture by \citet{evan88}, the ratio of \ion{C}{4} from our fit to the total [\ion{O}{3}] flux is $\sim$0.5.
In comparison, measurements of NGC 5548 yield a \ion{C}{4} : [\ion{O}{3}] ratio of $\approx$ 1.2, obtained
with 1$\arcsec$ (\ion{C}{4}) and 4$\arcsec$$\times$10$\arcsec$ ([\ion{O}{3}]) apertures \citep{krae98b}.
Based on this, the \ion{C}{4} flux from our derived profile is reasonable.
In contrast, in a sample of more luminous radio-loud AGNs, \citet{will93} detected no 
UV NLR lines, using the observed [\ion{O}{3}] line as a template, with upper limits on the 
\ion{C}{4} : [\ion{O}{3}] flux ratios of $\approx$ 0.1 -- 0.5.

\subsubsection{Covering Factor Model}

   For absorption features imprinted on multiple discrete background emission
sources, the normalized flux for the j$^{th}$ line can be written:
\begin{equation}
I_j = \Sigma_i [ R_j^i (C_j^i e^{-\tau_j} + 1 - C_j^i) ],
\end{equation}
where the i$^{th}$ individual emission source has fractional contribution to the total intrinsic (i.e.,
unabsorbed) flux, $R_j^i = F_j^i / \Sigma_i[F_j^i]$, and line-of-sight covering factor, $C_j^i$
\citep{gabe05}.
The effective covering factor associated with each line is the weighted combination
of the individual covering factors:
\begin{equation}
C_j = \Sigma_i [R_j^i C_j^i].
\end{equation}
These equations are extensions of the expressions given in \citet{gang99} for the
continuum and BLR to include an arbitrary number of background emission sources.
Combining equations 2 and 3 and solving for $\tau$ gives the familiar expression for
optical depth \citep{hama97},
\begin{equation}
\tau_j = \ln(\frac{C_j}{I - 1 + C_j}).
\end{equation}
The total column densities for each line are then obtained by integrating
\begin{equation}
N_j(v)=\frac{m_e c}{\pi e^2 f \lambda} \tau_j(v),
\end{equation}
over radial velocity \citep{sava91}, where the optical depths are derived
in each velocity bin from equations 3 and 4.

  Analysis of the Lyman lines in Paper II revealed the emission-lines are
only partially occulted by the UV absorbers.
Since the NLR is generally much more extended than the BLR and continuum source in AGNs,
we assume here a 3-component geometrical model in which the NLR is entirely unocculted
by all components of UV absorption.
Later, we explore this assumption based on geometrical constraints from our analysis.  
In Paper II, the emission-line and continuum covering factors were separated
assuming a single emission-line region.
Here, we incorporate the NLR -- BLR separation for our 3-component coverage model 
(hereafter 3-$C$) by solving for the continuum and BLR covering factor profiles, $C^c$ and $C^{BLR}$,
using the Lyman lines as in Paper II after first subtracting
a model of the Ly$\alpha$ NLR.
Given the mismatch between the Ly$\alpha$ NLR residual and \ion{C}{4} template
(see discussion above and Figure 3b), we fit the Ly$\alpha$ NLR with a narrower 
Gaussian that matched both the uncontaminated red wing and core, and
was constrained to be below the residual flux in the component~2 absorption feature 
($\sigma =$350~km~s$^{-1}$).  
Given the uncertainty in the NLR profile, we
cannot rule out that the deep component~2 feature partially occults the NLR of Ly$\alpha$.

    Using equation 3, we calculated effective covering factors for each absorption line,
using the derived $C^c$ and $C^{BLR}$ profiles, and $C^{NLR}=$0 at all radial velocities.
Figure 4 shows the resulting normalized unocculted flux levels (1 $- C$) for this 3-$C$
model compared to the observed normalized absorption profiles.
Regions where the residual fluxes in the absorption troughs are close to the unocculted
flux levels ($I \approx 1 - C$) indicate the lines are near saturation for this model.
For each line, results for both low and high-state spectra are shown to demonstrate
the variability and implications of the covering factor on observed line strengths.
To increase the S/N in each spectrum, we averaged multiple observations for each state,
selecting observations that exhibited the largest variations in absorption strengths and continuum
flux during the intensive monitoring.  
For the STIS lines, the low state shown is the mean of S5 -- S8, and the high state S10 and S12,
except for \ion{Si}{4}, which had the weakest absorption in observations S15, S16.
For lines in the {\it FUSE} spectrum, the low-state is the mean of F2 -- F4, and the high-state F5 -- F6.
Column densities for the low and high-states in each component computed with equations 4 and 5 
are given in Table 2.
Below, we describe the absorption lines for each kinematic component.

   For the \ion{N}{5} absorption, both doublet members are free of contamination over most of
the absorption profiles, providing a comparison between the doublet solution to the
covering factor ($C_d$) and our 3-$C$ model to test the effect of an unocculted NLR.
The unocculted flux levels for the doublet solution (1$-C_d$) 
are shown in Figure 4a with horizontal tick marks for \ion{N}{5}.
These values were derived in the cores of each kinematic component in the merged spectrum (Paper II).
Close comparison reveals important differences in the implied column
densities between the two covering factor models.
In components 1 and 3, the doublet solution gives unsaturated absorption.
This is seen in Figure 4a, where $I > 1 - C_d$ in the weaker doublet member 
(\ion{N}{5}~$\lambda$1242).  The resulting \ion{N}{5} column densities are
$\sim$ 8 $\times$ 10$^{14}$ cm$^{-2}$ (component~1) and 1.5 $\times$ 10$^{15}$ cm$^{-2}$ (component~3),
with equal values in the low and high-state spectra within measurement uncertainties.
Thus, the doublet solution implies the \ion{N}{5} absorption in components 1 and 3 is
unsaturated, with the above column densities, and without any detectable variability between flux states.
A similar result is found consistently for all STIS observations.
In contrast, the 3-$C$ coverage model implies \ion{N}{5} is near saturation in these
components, in both flux states, since $I \approx 1 - C$ consistently for both doublet lines.
The primary difference is due to the NLR flux underlying the red doublet members
of these components (see Figure 3b).
The lack of substantial variability in the equivalent widths (Figure 2a and 2c)
provides independent evidence for saturation of \ion{N}{5} in these components.
\ion{N}{5} in component~2 is unsaturated for both covering factor models, consistent with
the variability detected in its equivalent width.

\subsubsection{Component~1}

   Component~1, at $v_r =-$1350 km~s$^{-1}$, exhibits a rich absorption spectrum.
In addition to the common \ion{C}{4}, \ion{N}{5}, and \ion{O}{6} doublets, it
includes lines from the relatively low-ionization species \ion{Si}{4} and \ion{C}{2},
\ion{P}{5}~$\lambda$1118 (the \ion{P}{5}~$\lambda$1128 doublet line is contaminated 
with Galactic absorption and unmeasurable) and the metastable
\ion{C}{3}*~$\lambda$1175 complex (Paper II).
None of these lines is detectable in the other kinematic components in NGC 3783, and they are
only seen in a small fraction of intrinsic absorbers in AGNs (NGC 4151 is the only other
Seyfert with reported 
\ion{C}{3}* and \ion{P}{5} absorption, Bromage et al. 1985, Espey et al. 1998;
\ion{Si}{4} appears in $\sim$ 40\% of Seyfert absorbers, Crenshaw et al. 1999).
The \ion{C}{3}*~$\lambda$1175 absorption gives a tight constraint on the number density of
the absorber, as shown in \S 4.
\ion{H}{1} is detected in the Lyman series up to Ly$\theta$, while contamination
from Galactic absorption prevents detection of the higher order lines; thus the \ion{H}{1}
column density listed in Table~2 should be considered a lower limit.\footnote{In Paper II, 
we missed detection of Ly$\theta$ (as well as Ly$\zeta$) in component~1; 
thus the value quoted there, which was measured from the Ly$\epsilon$ line,
is smaller than the present study.}

   There is evidence that multiple, physically distinct absorption regions
are overlapping in kinematic component~1 (KC01; Paper II).
In Paper II, it was shown the \ion{Si}{4} covering factor from the doublet solution is
lower than that derived from the Lyman lines and the \ion{N}{5} doublet.
This is confirmed for the new 3-$C$ model:
the column density measured for the red \ion{Si}{4} doublet line with this covering
factor is three times greater
than the blue line, indicating the actual \ion{Si}{4} covering factor is significantly lower.
In contrast, the 3-$C$ model gives consistent results
for the two \ion{N}{5} doublet members:  saturation in the blue-wing and core
in both flux states (see above discussion), and similar column densities in the red-wing
of the profile where the absorption depths diverge from the unocculted
flux levels (i.e., the absorption is unsaturated).
Similarly, \ion{C}{4}~$\lambda$1548 and \ion{O}{6}~$\lambda$1038 are consistent with being saturated
in their blue wings and cores with the 3-$C$ model (the other doublet members for these ions are
unmeasurable due to contamination with other absorption), and
\ion{C}{4} diverges substantially in the red wing (\ion{O}{6}~$\lambda$1038 is contaminated with a
detector artifact at these velocities and cannot be tested).
Additionally, there are ion-dependent structural differences in the absorption profiles.
As described in Paper II, the red wings of some lines, particularly \ion{N}{5} and Ly$\alpha$, extend to
lower velocities than \ion{Si}{4} and \ion{P}{5}.  Interestingly, it is at these velocities
that the \ion{N}{5} and \ion{C}{4} profiles diverge from the unocculted flux levels from the covering
factor model.

  Thus, we treat the absorption from this kinematic region as coming from
two physical components:
component~1a has relatively low covering factor and low-ionization and
gives \ion{Si}{4}, \ion{C}{2}, \ion{C}{3}*, and \ion{P}{5};
component~1b is more highly ionized and has higher covering factor, contributing
strongly to \ion{C}{4}, \ion{N}{5}, \ion{O}{6} and the Lyman lines.
The latter lines will have contributions from both components, with absorption from
component~1a buried in the higher covering factor absorber \citep[e.g., see][]{krae03}.
To measure the \ion{Si}{4} column density, we used the covering factor derived from the
doublet pair ($C=$0.35; Paper II).
For the other lines associated with component~1a, there is no independent
measure of the covering factors (it is not possible to separate the emission-line
and continuum covering factors for this component from the single \ion{Si}{4} doublet).
We assumed $C =$1 for these lines because their
underlying emission is predominantly continuum flux.
Since they are all weak (see Figure 4), the measurements will not be too far off
unless $C^c$ is very small in this component.
For the \ion{O}{6}, \ion{N}{5}, and \ion{C}{4} lines, we adopted the 3-$C$ model.
Since all these lines show saturation in this component over much of the profiles,
we derived lower limits on their integrated column densities by adding estimated
uncertainties to the normalized fluxes, $I + \Delta I$, in equation 4.
Measured column densities and limits for the low and high-states are listed
in Table 2.
For modeling purposes (\S 5.2), we also measured the \ion{C}{4} and \ion{N}{5} absorption in the
red wing of the profile, where the lines are unsaturated and there is no contribution
from component~1a based on the \ion{Si}{4} profile ($v_r = -$1270 --  $-$1170 km~s$^{-1}$).
This gives 3.7$\times$10$^{14}$ and 1.0$\times$10$^{14}$~cm$^{-2}$
for the \ion{N}{5} and \ion{C}{4} column densities, respectively.

  As described in \S 3.1, the \ion{Si}{4} absorption varied, with a general inverse correlation
with continuum flux.
With the relatively low \ion{Si}{4} covering factor, the moderate equivalent
width variations translate into peak changes of a factor of $\sim$ 4 between the mean
low and high states during the intensive phase defined above.
The column density decreased somewhat less in the high-state 
S10 and S12 observations relative to the mean low-state (a factor $\sim$ 2).
Figure 4c shows the same trend for \ion{C}{2}: weak absorption is detectable in
the ground state (\ion{C}{2}~$\lambda$1335) and fine-structure
(\ion{C}{2}*~$\lambda$1336) lines in the low-state spectrum, but was not
distinguishable from noise in the high-state.
The weak \ion{P}{5}~$\lambda$1118 and \ion{C}{3}*~$\lambda$1175 absorption appears to decrease
as well, with only $\sim$1.5$\sigma$ detections in the high-state, though
the limited S/N in the spectra of these lines
makes this less certain.  No significant variability was detected in Ly$\theta$, 
nor any other Lyman series lines, between flux states.

\subsubsection{Component~2}

  In component~2 ($v_r =-$550 km~s$^{-1}$), the \ion{N}{5} and \ion{C}{4} absorption is relatively weak and
variable (Figures 2b and 4a).
Our adopted covering factor model gives unsaturated absorption in these doublets,
consistent with the observed variability.
In contrast, Figure 4b shows the \ion{O}{6}~$\lambda$1038 line is strong and
matches the model unocculted flux levels
nearly identically across the entire profile in both states, consistent with
heavy saturation of this line.
Lyman line absorption is measurable only up to Ly$\gamma$ in component~2 due to
strong contamination of the higher order lines with Galactic absorption.
The Ly$\gamma$ line does not vary significantly, thus we adopt the \ion{H}{1} column density measured
from this line as a lower limit.
Figure 4b shows both the \ion{C}{3}~$\lambda$977 and \ion{N}{3}*~$\lambda$991 lines
are contaminated with moderate Galactic H$_2$ absorption at component~2 velocities, with neither line
showing strong evidence for the presence of intrinsic absorption.  Upper limits on these
lines measured after the removal of the H$_2$ model are given in Table~2.

\subsubsection{Component~3}

   In component~3 ($v_r =-$725 km~s$^{-1}$), the primary constraints for modeling are the 
\ion{C}{3}~$\lambda$977 line and
the \ion{C}{4} doublet.  Both lines exhibit variability, decreasing in epochs with higher
continuum flux.
The \ion{C}{3} line is relatively strong in the low-state {\it FUSE} spectrum, but
is similar to the noise level in the high state (Figure 4b), giving only an upper limit.
Figure 4a shows that for the adopted covering factor model, the \ion{C}{4}~$\lambda$1551 line
is near saturation in the low-state, but well above the unocculted flux level in the high-state.
Thus, for this model, the equivalent widths in \ion{C}{4}
vary only moderately (Figure 2c), but the variation in column density is large between flux
states due to the low effective covering factor ($C \approx$ 0.35 in the core of component~3).
The \ion{C}{4}~$\lambda$1548 line is blended with the component~1 \ion{C}{4}~$\lambda$1551 line,
thus no comparison of our adopted model with the doublet solution is possible.
The \ion{O}{6}~$\lambda$1038 and \ion{N}{5} doublet lines (see above) are consistent
with being saturated in both flux states, based on the unocculted flux levels.
The highest order Lyman line not contaminated with other absorption in component~3 is Ly$\epsilon$.
Since it does not vary significantly between flux states, we adopt the resulting
\ion{H}{1} column density from this line as a lower limit.
The \ion{N}{3}~$\lambda$989 component~3 line coincides with the strong \ion{N}{3}*~$\lambda$991 
component~1 line, and cannot be
measured. \ion{N}{3}*~$\lambda$991 in component~3 is moderately contaminated with Galactic H$_2$.
Given the limited S/N in this region, we take the measurement of \ion{N}{3}* after the 
removal of the H$_2$ model as an upper limit.

\section{Constraints on the Density from the Metastable \ion{C}{3} Absorption}

   The \ion{C}{3}*~$\lambda$1175 multiplet lines have been used as a
density diagnostic for AGN absorbers in several studies. However, as pointed
out by \citet{beha03}, the high densities derived in these studies, including our Paper II,
were based on calculations of level populations that only treated the
$^3$P$_1$ level.
The $J=$0 and 2 levels have much lower radiative transition probabilities
to the ground state ($A \approx$0 and 5$\times$10$^{-3}$ s$^{-1}$, respectively) than
the $J=$1 level ($A =$75 s$^{-1}$), and thus are populated at
densities that are lower by several orders of magnitude
\citep[see Figure 1 in][]{bhat93,kast92}.
This has important consequences for the interpretation of the absorbers
since the gas density is needed to determine the location and physical depth of the gas.
To address this, we computed the relative populations of the $^3$P$_J$ levels for a range of
densities and electron temperatures, $T_e$, that are expected for the photoionized AGN outflows
seen in the UV.  Our calculations extend the results of Bhatia \& Kastner (1993), who give
results applicable to collisionally ionized plasmas with $T_e \geq$ 40000 K, to lower temperatures.
Collisional excitation and de-excitation and radiative decay
between all levels were treated. We treated the 6 lowest terms/levels of the $C^{+2}$
ion: the ground state 2s$^2$ $^1$S$_0$ term, the three 2s2p $^3$P levels, 2s2p $^1$P$_1$,
and 2p$^2$ $^3$P.
Temperature dependent collision strengths from \citet{berr85,berr89} were used for
transitions between levels and terms, respectively, where we have interpolated between their
listed temperatures.  Radiative transition rates were obtained from \citet{bhat93}, \citet{mort91}, 
and the NIST Atomic Spectra Database website.

    Figure 5 (top panel) shows the computed populations over a large range in density for $T_e=$
16000, 20000, and 40000 K.
This shows that the \ion{C}{3}*~$\lambda$1175 absorption complex
can serve as a powerful probe of the physical conditions in the absorber.
The {\it relative} populations of the three levels are very sensitive
to the electron density, but insensitive to temperature.
This is seen in the bottom panel of Figure 5, where the ratio of the $J=$2 : $J=$0 level populations are
plotted for the temperatures shown in the top panel.
Additionally, the absolute populations of the $^3$P levels are very sensitive to the
gas temperature.

   In practice, full utilization of these lines as diagnostics
requires that the absorption features are sufficiently narrow so that the
individual lines in the complex are not heavily blended, and sufficient
resolving power to separate the lines.
These conditions are met with the STIS spectra of component~1 in NGC 3783 as seen
in Figure 6, which shows only mild blending of the individual lines.  The location of the six multiplet
lines are marked and identified by the $J$ level of the transition.
A fit to the \ion{C}{3}*~$\lambda$1175 complex is shown as a dashed line, using
the width and centroid of the \ion{Si}{4} absorption profile; best-fit column densities
for each level are given below the spectrum.
The column density ratio of the $J=$2 : $J=$0 levels, $N_{J=2} / N_{J=0} =$ 2.9$\pm$1.4,
gives $n_e$=3$^{+5}_{-1.5}$$\times$10$^4$~cm$^{-3}$, independent of temperature.
The lack of detection of the $\lambda$1174.93 line is consistent with this density since
Figure 5 shows the population of the $J=1$ level will be approximately four orders of magnitude
lower than the other levels.
Use of these level populations as a temperature diagnostic requires knowing the total
abundance of the C$^{+2}$ ion; since the \ion{C}{3}~$\lambda$977 line is unmeasurable
due to contamination (and saturated for the implied column density),
this requires the results of photoionization modeling, which is
presented in \S 5.

    Our calculations do not treat the effects of the radiation field on the level/term
populations. This could decrease the ground state or metastable level
populations due to continuum pumping, or increase the relative populations of the metastable levels
due to recombination followed by cascade to the 2s2p levels.
We have computed photoionization models with Cloudy to test this and find it has a negligible
effect on the metastable level populations in this component.

\section{Photoionization Models}

\subsection{Input Parameters and Assumptions}

   We compare the observed ionic column densities with
predictions from the photoionization modeling code Cloudy \citep{ferl98} 
to constrain the physical conditions and location of the intrinsic absorbers.
The absorbers are assumed to be uniform plane parallel slabs of constant density, 
photoionized by the AGN at a distance $R$ from the central source.
These calculations apply for a gas in ionization
and thermal equilibrium with the ionizing flux; this is explored in \S 5.3 
based on the observed variability in the absorption.
The models are specified by the spectral energy distribution (SED) of the ionizing
continuum, the total hydrogen column density ($N_H$) and
elemental abundances in the absorber, and the ionization parameter 
($U= Q/4 \pi R^2 n_H c$), which gives the ratio of the density of H ionizing photons 
at the face of the absorber to the gas density, $n_H$.

  The choice of SED for the unobservable ionizing continuum 
is a source of modeling uncertainty in AGNs \citep[e.g.][]{math94,kasp01}.  
To facilitate the comparison with the earlier UV results, we adopted an SED based on the
one used by KC01.
This consists of multiple power-law components ($F_{\nu} \propto \nu^{-\alpha}$)
constrained by the UV and X-ray observations.
The UV power-law component is based on continuum-flux measurements
in the two epochs with simultaneous STIS and {\it FUSE} observations. 
After correcting the spectrum for Galactic extinction \citep[$E(B-V) =$ 0.119; ][]{reic94} 
with the \citet{card89} reddening 
curve, the continuum flux ratio measured at 950 \AA~and 
1690 \AA~gives a spectral index of $\alpha_{\nu}^{UV} \approx$~1. 
As discussed in KC01, extrapolating the UV power-law
to X-ray energies far overestimates the observed flux, thereby requiring a spectral
break in the unobservable EUV -- soft X-ray flux to connect
the absorption in the two bands.
Thus, we extrapolated from the observed UV flux at the
Lyman limit to the observed flux at 0.6 keV in the {\it Chandra} spectrum, giving
a spectral index $\alpha_{\nu}^{EUV}=$~1.4.
For the hard X-ray index, we used the value derived from the merged
{\it Chandra} spectrum in Paper I, $\alpha_{\nu}^{X} \approx$ 0.7 ($\Gamma =$1.7).
For this SED and the adopted distance 
of 39~Mpc for NGC~3783 \citep[using $z =$0.00976 from][and $H_o =$75~km~s$^{-1}$]{deva91}, 
this gives an H-ionizing photon luminosity 
$Q =$1.8$\times$10$^{54}$~s$^{-1}$.
The SED adopted here is somewhat different from the one used to model the X-ray absorption 
in Paper~IV, which had $\alpha =$0.5, 4.7, and 0.77 in the energy ranges 0.002 -- 0.04, 
0.04 -- 0.1, and 0.1 -- 50 keV).
We recomputed our models derived in the next section with the SED used in Paper~IV
and find no significant differences in the predicted column densities for the ions measured 
in the UV spectrum, after
scaling the ionization parameter to account for the somewhat larger relative EUV
flux used in the X-ray models.
The issue of SED is addressed further below in the discussion of the combined UV and X-ray
modeling results (\S 6.2).
We adopted the roughly solar elemental abundances \citep{grev89} used in the KC01 models, 
and assumed no dust is present in the absorber.

    The combined UV and X-ray spectrum of NGC 3783 shows that a large range of 
ionization states is present in all kinematic components and there are multiple zones 
of ionization overlapping at all absorption velocities (KC01; Paper IV).
Thus some lines, particularly from more highly ionized species, 
may be comprised of blends of different physical components having 
different covering factors and optical depths; evidence for this 
was presented in \S 3 for kinematic component~1, for example.
Therefore, we take as primary constraints the 
lines least likely to be affected by blending or saturation.
These include lines from the lowest-ionization species detected in each kinematic
component, ions with low abundances of the parent element (e.g., \ion{P}{5}), and ions in
excited states (e.g. \ion{C}{3}*).
Additionally, upper limits on column densities from non-detections provide
unambiguous constraints.

\subsection{Model Results for Mean Low and High State Spectra}

    Figure 7a shows the solutions in log($U$) and log($N_H$) from a grid of 
photoionization models that match the measured column densities for component~1. 
The contours on the left are for the low covering factor absorber, component~1a.
The thickness of the contour for each ion spans the range of solutions corresponding 
to estimated uncertainties in the measured column densities \citep[e.g., see][]{arav01}.
Figure 7a includes solutions for \ion{Si}{4}, \ion{C}{2}, \ion{C}{3}, and \ion{P}{5} 
measured in the low-state spectrum, plotted with hatched marks.
The \ion{C}{3} solutions are for the measured metastable level column densities, 
using the electron density 
derived from the \ion{C}{3}*~$\lambda$1175 feature in \S 4 and an electron temperature  
of $T_e \approx $1.8$\times$10$^{4}$ K, consistent with the best-fit Cloudy model.
The solutions to the high-state column densities (mean of S15 and S16) 
for \ion{Si}{4} are also included on the plot, shown as the contour with no
hatch marks.
These were shifted along the x-axis by log($U$)$=-$0.23,
corresponding to the peak amplitude of flux variation observed during the intensive monitoring.
This gives the high-state solutions in terms of the low-state ionization parameters,
allowing a direct comparison of the two flux states on the same grid:
for a correct model with both states in ionization equilibrium and an 
ionization parameter that scales as the observed UV continuum flux, the low-state and
shifted high-state solution contours would overlap.
High-state solutions occupying regions in $U$ - $N_H$ above and to the 
left of the low-state solutions overestimate the variability observed between states, while 
solutions to the right and below underestimate the variability.

     Figure 7a shows all ions plotted for the low-state spectrum are fit well 
by an extended, narrow range of models in $U$ -- $N_H$ parameter space, with lower bounds 
$log(U) >$ $-$1.7 and $log(N_H) >$ 20.3.
For higher ionization models, these ionic abundances (particularly \ion{Si}{4} and \ion{P}{5}) 
are very sensitive to small changes in $U$ and $N_H$, as seen in the narrowing of the 
solutions spanning the measured limits on the column densities.
This is due to the \ion{He}{2} opacity -- these solutions are in the region in parameter space where 
the \ion{He}{2} edge becomes optically thick.
In these models, the relatively low-ionization species in component~1a
exist primarily in a small region in the back end of the slab, 
where the continuum flux is heavily filtered, thus their abundances are
greatly affected by the sensitivity of the \ion{He}{2} opacity in the Stromgren 
shell to the model parameters \citep[e.g.][]{krae02}.
For these high-ionization models, the range of solutions becomes linear in $U$ - $N_H$, tracing
the \ion{He}{2} column density contour.

     Combining the shifted solution to the high-state \ion{Si}{4} column density, the range
of solutions is limited to low-ionization values.
The overlap between \ion{Si}{4} high and low-states gives slightly lower $U$ than 
the fit to all low-state lines, but the solutions are close.  
For models with $log(U) > -$1.5 in the low-state, Figure 7a shows the  
\ion{Si}{4} variability is predicted to be significantly greater than observed.
This discrepancy becomes pronounced for higher-ionization solutions: e.g., a factor of 1.7 increase
in ionizing flux from the log$U=-$1, log$N_H=$21.6 model results in a \ion{Si}{4} column density of
$\sim$ 10$^{12}$ cm$^{-2}$, which is 40 times weaker than observed.
Thus, based on these models of the mean high and low-state spectra,
the lowest ionization absorber in component~1 has best-fit solution log($U$)~$= -$1.6$^{+0.2}_{-0.2}$, 
log($N_H$)~$=$~20.6$^{+0.4}_{-0.3}$.
The limits on the high-state \ion{C}{2}, \ion{C}{3}*, and \ion{P}{5} column densities 
(not shown on Figure 7a) are also compatible with these solutions.
We note the predicted column densities for \ion{O}{6}, \ion{N}{5}, and \ion{C}{4} 
in this absorber would produce heavily saturated absorption in their resonance doublet lines,
with $N_{ion} >$ 5$\times$10$^{15}$ cm$^{-2}$ for each ion.
Model predictions for all ions are listed below their measured values in Table 2.   
   
    The physical parameters of the absorber with higher covering factor, component~1b,
can also be constrained. 
For optically thin conditions, the \ion{C}{4} and \ion{N}{5} column densities each scale
linearly with $N_H$, but have different dependences on $U$; thus, the \ion{C}{4}~:~\ion{N}{5} 
ratio uniquely determines the ionization parameter, independent of the total column density.
As a result, assuming the ionization structure in component~1b is uniform over all radial velocities, 
the \ion{C}{4} and \ion{N}{5} column densities measured in the unsaturated red wing 
determines $U$ for the entire absorber.
Our measurements give log($U$)$= -$0.4$^{+0.3}_{-0.2}$.
A lower limit on the total column density comes from the lower 
limits measured on \ion{C}{4} and \ion{N}{5} over the full, saturated profile.  For the above $U$, this gives
log($N_H$)~$\geq$~20.3.
The \ion{O}{7} column density measured in the {\it CXO} spectrum over the
velocity range coinciding with component~1 (see Figure 10 in Paper I) provides an upper limit on $N_H$.
Our fit to this high-velocity region of the profile gives $N_{O VII} =$~5$\times$10$^{16}$~cm$^{-2}$,
with an upper limit of $\leq$~10$^{18}$~cm$^{-2}$ at the 90\% confidence level.
Incorporating this as an upper limit on \ion{O}{7} gives $N_H \leq$~21.4 for component~1b.
The region in $U$, $N_H$ parameter space spanned by these limits is also plotted on Figure 7a.

     The modeling constraints for component~2 are \ion{C}{4} and \ion{N}{5} in both the low and high states, 
and the upper and lower limits on \ion{C}{3} and \ion{H}{1}, respectively.
These solutions are shown on the $U - N_H$ plot in Figure 7b; \ion{N}{5} and \ion{C}{4} are
shown as hatched contours in the low-state and as unfilled dashed contours in the high-state,
which are shifted to account for the flux difference between states as in Figure 7a.
In each state individually, the \ion{C}{4} and \ion{N}{5} solutions overlap over an extended  
range in $U - N_H$; however, the combined high and low-states  
are simultaneously fit by only a small region of parameter space selecting the lower 
ionization solutions; log($U$) $= -$0.45$^{+0.2}_{-0.1}$ (low-state values), 
log($N_H$) $=$ 20.4$^{+0.6}_{-0.1}$.
These solutions are also consistent with the \ion{H}{1} and \ion{C}{3} limits.

    In component~3, \ion{C}{3}~$\lambda$977 is clearly detectable in the low-state {\it FUSE} 
spectrum, representing the lowest-ionization species detectable in this component,
but has only an upper limit in the high state (see \S 3.2.5 and Figure 4b).
For our adopted effective covering factor, \ion{C}{4} is near saturation in the low-state, 
but well below the residual flux level in the high-state giving a correspondingly small 
column density.  The \ion{N}{5} doublet is consistent with being saturated in both states.
The model results are shown in Figure 7c, with high-state solutions shifted as in Figures 7a and 7b.
This shows the \ion{C}{4} high-state and \ion{C}{3} low-state column densities (hatched contours) 
are simultaneously matched by a 
restricted space of correlated $U$ and $N_H$ values, with log($U$)$\gtrsim -$1.2, 
log($N_H$)$\gtrsim$19.
Including lower limits on \ion{N}{5} and \ion{H}{1} places lower limits of 
log($U$)~$\gtrsim -$0.7 and $\gtrsim -$0.6, respectively.
An upper bound on the solution comes if we require the model X-ray columns do
not exceed those measured in the {\it CXO} spectrum.
The \ion{O}{7} column density integrated over all radial velocities 
($N_{O VII} =$~10$^{18}$~cm$^{-2}$ from Paper I), provides the most 
rigid constraint and is plotted in Figure~7c.
Incorporating this limit and assuming all the lines measured in the UV
arise in a single physical component gives log($U$)$= -$0.5$^{+0.1}_{-0.1}$,
log($N_H$)$=$ 21.1$^{+0.3}_{-0.2}$.

    These modeling results can be compared to those from KC01, which were based on the first
STIS observation.  The ionization and total column density for component~1a are substantially
greater here than in KC01. This is due to the constraints implied by the detection of 
\ion{C}{2} and \ion{P}{5} in the data presented in this study, and the use of a lower covering 
factor for \ion{Si}{4}, giving a larger \ion{Si}{4} column density.
A reliable covering factor for \ion{Si}{4} was made possible by the high S/N in the merged STIS 
spectrum (Paper II).  
The solutions to components 2 and 3 presented above have similar total column densities
as the KC01 solutions, but have somewhat lower ionization parameter ($\sim$ 0.3 -- 0.4 dex).  
For component~2, this lower ionization was imposed on the solution primarily by
the magnitude of variability observed in \ion{C}{4} and \ion{N}{5} between high and low flux states. 
For component~3, the detection of \ion{C}{3}~$\lambda$977 
in the {\it FUSE} spectrum and the limit placed by \ion{O}{7} measured in the {\it CXO} spectrum
drove the solution to lower ionization.

\subsection{Time-Dependent Ionization Solutions}

   Here we explore variability timescales for the UV absorbers.
With the density constraint derived for component~1a from \ion{C}{3}*, 
the time-dependent ionic populations for this absorber can be probed in detail
for comparison with the observed variability, thereby providing better
constraints on its physical conditions. 
For the other absorption components, the observed magnitude and timescales for variability
in ionic populations in response to the continuum flux variations can be used to
place limits on the density. 

\subsubsection{Detailed Variability Calculations for Component~1}

    The response time for an ion is a strong function of several factors:  its population ($n_i$) relative 
to adjacent ionization stages, the electron density, and the magnitude of change in ionizing flux incident 
on the absorber\citep[e.g.][]{hama97}.  We computed time-dependent ionic populations in response to changes in the ionizing continuum  
for the gas in component~1 based on the density and ionization solutions determined above.
The problem involves solving for ionic abundances, $n_i$, from the system of first-order
differential equations:
\begin{equation}
\frac{dn_i}{dt} = - [p_i + n_e \alpha_{rec,i-1}(T)] n_i + n_e n_{i+1} \alpha_{rec,i}(T) + p_{i-1} n_{i-1},
\end{equation}
where $p_i$ and $n_e \alpha_{rec,i}$ are the ionization rates from and recombination rates to ionization 
stage $i$ (e.g., Krolik \& Kriss 1995).
Output from the low and high-state Cloudy equilibrium models were used for the parameters in equation 6, 
corresponding to initial and final states associated with a change in ionizing flux.
The time-dependent ionic abundances were then solved using the Runge-Kutta-Fehlberg method, which compares 
fourth and fifth-order Runge-Kutta estimates to adjust the step size.
The rates include all ionization and recombination processes treated by Cloudy \citep[see][]{ferl98}.
We consider simple step-function increases and decreases in the ionizing flux below.

    For example, to compute the time-dependent component~1 \ion{Si}{4} column density in response to an increase 
in ionization from the low to high states, 
the initial values of all ionic species of silicon ($n_i (t=0)$) were set to the 
values from the low-state equilibrium model (log($U$)$ =-$1.6, log($N_H$)$=$20.6).
The ionization rates, $p_i$, were set to the values from the corresponding high-state equilibrium 
model (log($U$)$=-$1.37).  This assumes the ionization rates change instantaneously, which is valid
if they scale linearly with the ionizing flux, as is the case in the model considered here.
The recombination rates differ somewhat between the initial and final state equilibrium models
because of their temperature dependence.
However, for the time intervals and magnitude of flux variations considered here, the absorber is not in
thermal equilibrium, as determined by the thermal timescale: 
\begin{equation}
t_{thermal} = \frac{3/2 n_e k T}{n_e^2 \Lambda (T)},
\end{equation}
where the numerator is the total thermal energy and the denominator
the net cooling rate in the gas per unit volume, with $\Lambda (T)$ representing the difference 
between total heating and cooling rates, which are obtained from the Cloudy calculations.
Even for extreme changes in ionizing flux, the thermal timescale (taken here as the e-folding time) 
is of order 1/2 year for these conditions.
It is much longer than this for the relatively minor flux changes observed in NGC 3783, due to the 
small value of the net cooling rate.
Thus, the gas is not in thermal equilibrium and the actual value of the temperature reflects an 
average of the long-term history of the cooling and heating rates as they respond to variations in the ionizing flux.
For the conditions considered here, the temperature can be assumed to be constant. Thus, we adopt 
the recombination
rates from the initial state models in our solutions to equations 6.

   Figure 8 illustrates the dependence of time-dependent populations 
on the amplitude of ionizing flux variations.
The time-dependent behavior of the \ion{Si}{4} column density is given for both increases and decreases
in ionizing flux by factors 1.4, 1.7, and 3, showing the time needed to achieve a given level of
variability is sensitive to the amplitude of flux change.
For models with increased ionization, we used the low-state model solution as the initial state, 
and for decreased ionizing flux, the high-state model.
For large decreases in ionizing flux, the recombination timescale (i.e., e-folding time) is seen to approach the
expression from \citet{krol95}, $t_{recomb} = n_i / (n_{i+1} n_e \alpha_{rec,i})$;
$t_{recomb} =$ 5 days for \ion{Si}{4} for the model considered here.  
Figure 8 also shows the time interval corresponding to a given fractional change 
in the ionic abundance is similar for cases of {\it ionization} and {\it recombination}   
for the moderate flux variations considered here.  
However, for very large flux changes, our calculations show the ionization
timescale becomes much shorter than the recombination timescale.

    We calculated the detailed variability for the component~1a absorber based on 
the observed flux variations in individual STIS observations.
In Figure 9, calculations for multiple step-function variations in flux, corresponding
to the continuum light curve during our intensive monitoring, are compared with the
measured \ion{Si}{4} column densities.
The solid line shows results for the model derived in \S 5.2 based on the 
low and high-state solutions; the UV light curve is shown in the bottom panel for comparison.
The rates and initial ionic populations were taken from the low-state Cloudy model, log($U$)=$-$1.6,
log$N_H$=20.6.
At each epoch, the ionization rates were then scaled by the change in flux observed in 
the subsequent observation, and the time-dependent populations computed using equation 6.
The overall \ion{Si}{4} variability is seen to be matched well by this model.
It reproduces the somewhat damped decrease in \ion{Si}{4} column density observed in
high-states S10, S12 following the increased flux in those observations, 
and the further decrease in later epochs, where the absorption
is a minimum at S15.

     In \S 5.2, an extended range of solutions, with correlated $U$ and $N_H$ values, 
was found to match the low-state column densities in component~1a reasonably well, 
with the high-state solution selecting the lower $U$ models (Figure 7a).
If the physical conditions in this absorber are such that the \ion{Si}{4} population is 
not able to respond to the ionizing continuum sufficiently, the gas could in principle be in a 
higher ionization state than modeled by assuming full equilibration between states.
With the density for component~1a determined independently from \ion{C}{3}*, 
it is possible to test this directly.  Thus, we computed the time-dependent \ion{Si}{4} 
population for higher $U$ models to test its variability.
Figure 9 shows results for a model with log($U$)~$= -$1, log($N_H$)~$=$21.5 (dashed line), 
which was selected from the low-state solutions in Figure 7a.
This high $U$ model is seen to far overestimate the variability in \ion{Si}{4},
and thus is excluded by our timing analysis.

\subsubsection{Constraints on Density in Components 2 and 3 from
Observed Variability}

    Components 2 and 3 do not have direct estimates of the density,
and thus are less well constrained than component~1.
However, lower limits on their densities can be derived based on the
observed variability.
This constraint comes from the characteristic time separation between the low and
high-state epochs,
which provides an upper limit on the timescale for changes in ionic abundances
as they respond to the ionizing continuum.

      For component~2, we computed the time-dependent \ion{C}{4} and \ion{N}{5}
abundances in the same manner as above for component~1 to derive these limits.
We assumed a simple step-function change in flux of a factor 1.7, 
corresponding to the continuum variation between the mean low and high-states
during the intensive monitoring.
We parameterized the observed variability for these models 
by requiring the ionic populations varied by at least
half the amount observed over a characteristic timescale, 
which we define as the time interval between the mean of the 
low states (S5 -- S8) and mean of the high states (S10, S12), 
$\sim$ 20 days, where maximum change in these column densities occurred.
This places a lower limit on $n_e$.
The e-folding variability times are proportional to the inverse 
of the density ($t \propto n_e^{-1}$).
Thus, we computed the model for a fiducial
density and then solved for the value of $n_e$ that reproduces the
required level of variability over the characteristic 
timescale.

    To demonstrate this, Figure 10 shows the time-dependent
\ion{N}{5} and \ion{C}{4} ionic abundances for component~2
based on the solution in \S 5.2 (log($U$)=$-$0.45, log($N_H$)=20.4).
These were computed from equation 6 for a fiducial density of 
$n_e=$10$^4$ cm$^{-3}$, thereby 
giving $N_{ion} / N_{ion, t=0}$ as a function of $t \times (n_e/10^4)$.
Horizontal dashed lines mark the 50\% level of variability;
dotted lines give the final, equilibrium values.
The lower limits on density follow straightforwardly from 
the intersection of the abundance curves with the variability
limits (dashed lines), which is assumed to occur over a 20 day interval:
$n_e = t/20 \times 10^4$~cm$^{-3}$.
The \ion{N}{5} variability gives $n_e \geq$ 1.3$\times$10$^3$~cm$^{-3}$
and \ion{C}{4} gives $n_e \geq$2.3$\times$10$^3$~cm$^{-3}$ for component~2.
Component~3 is less well constrained because for each of the line constraints,
at least one state provides only a limit on the column density.  A similar analysis 
for this component gives $n_e \geq$7.5$\times$10$^2$~cm$^{-3}$.

\section{Interpretation}

\subsection{Constraints on Physical Conditions and Geometry of the UV Absorbers}

  We now derive constraints on the physical conditions and
geometry of the absorbers based on our above analysis.
The best constraints are for the high-velocity outflow region,
component~1, due to the density measured directly from
the metastable \ion{C}{3}* feature (\S 4).
From the expression for the ionization parameter given in \S 5.1,
the distance between the absorber and central source 
can be solved for the low-ionization absorber (component~1a),
using the derived values of the number density ($n_H \approx n_e$/1.2),
$Q$, and $U$ from the modeling in \S 5.2.  This gives $R =$7.7$^{+3.1}_{-4.6}\times$10$^{19}$~cm (25 pc).
If the individual kinematic components are uniform clouds, i.e., having {\it internal}
volume filling factors of unity, then their radial physical depths 
can be determined straightforwardly from the ratios of the total column
density to the number density, giving $\Delta R=$1.3$\times$10$^{16}$~cm for
component 1a.  Results are summarized in Table 3.

  In Paper III, we showed the radial velocities for all component~1
lines in the STIS spectra decreased at a rate
of $\sim$~50~km~s$^{-1}$~yr$^{-1}$. This includes \ion{Si}{4}, which comes
from the low-ionization, low-covering subcomponent, and the \ion{N}{5} and
\ion{C}{4} lines which are predominately from component~1b.
Thus, these absorbers appear to be dynamically linked and, hence, co-located.
Using this result, and adopting the ionization parameter for component~1b
derived in the unsaturated red-wing of \ion{C}{4} and \ion{N}{5} (\S 5.2),
the density for this higher ionization component can be solved from:
$n_{H,1b} = n_{H,1a} \times U_{1a} / U_{1b}$, giving
$n_{H,1b} =$ 1.9$\times$10$^3$~cm$^{-3}$.
A broad range of values for the radial depth for component~1b follow from
the loose constraints on $N_H$,
8$\times$10$^{16}$$\leq \Delta R \leq$1$\times$10$^{18}$~cm.
This can be compared to the lower limit on the projected transverse size
of this absorber, determined by the partial coverage of the BLR from
the Lyman line analysis and the size of the BLR based on reverberation
mapping (Onken \& Peterson 2002), giving $X_T \geq$ 1$\times$10$^{16}$~cm
(Paper II).  The transverse size of component~1a is not well constrained because
the details of the covering factors of the individual sources are
not known; the only constraint is that the UV continuum source is at
least partially covered by this absorber.

  For components 2 and 3, the constraints derived on $n_e$ from the observed  
variability (\S 5.3.2) give upper limits on their distances from the
AGN and radial depths.
Our derived lower-limit on $n_e$ in component~2 based on the calculated
response times for \ion{C}{4} and \ion{N}{5} gives $R \leq$~7.4$\times$10$^{19}$~cm (24 pc) and
$\Delta R \leq$~10$^{17}$~cm.
The density limit for component~3 implies $R \leq$~1.4$\times$10$^{20}$~cm (45 pc) and
$\Delta R \leq$1.3$\times$10$^{18}$~cm.
Lower limits on the projected transverse dimensions of these absorbers follow from
the BLR covering factors, giving $X_T \geq$ 1 -- 2$\times$10$^{16}$~cm.
Noting that $\Delta R \propto n_H^{-1}$ and $R \propto n_H^{-1/2}$, 
the absorber's geometries are seen to depend strongly on their locations when
combined with the constraints on the transverse sizes.
For example, if the absorbers are uniform clouds having roughly spherical geometries,
with dimensions approximately equal to the lower limits given by the BLR covering
factors, then distances $\sim$~10 and 5~pc are implied for components 2 and
3, respectively.
Smaller distances would require flattened geometries. 
For example, if they are located at 1/2~pc,
then they would be extremely thin shell structures, having radial
dimensions at least 100 -- 500 times smaller than their transverse sizes.

   The results summarized in Table 3 show all UV kinematic components
have low {\it global} volume filling factors ($\Delta R/R \ll$~1), and are
consistent with being relatively small, discrete clumps.
Additionally their derived distances (or limits) are consistent with
the location of the inner narrow line region in AGNs, and inside the
more extended, diffuse NLR.
The geometry implied by these results is consistent with our earlier 
assumption that the NLR is unocculted by the individual components of 
UV absorption in deriving the covering factor model (\S 3.2.2).

\subsection{Connection between the UV and X-ray Absorption}

   Here, we explore the connection between the observed UV and X-ray absorption.
The absorption in the two bandpasses shows similar kinematic structure: 
in Papers I and II, all X-ray lines
having sufficiently high-resolution and S/N were found to span the radial velocities 
of the three UV kinematic components.  Modeling 
of the {\it CXO} spectrum in Paper IV showed the X-ray absorption is highly inhomogeneous, 
requiring three ionization zones that span a range $>$ 50 in $U$ to reproduce the full set 
of lines.  A summary of the modeling and geometric constraints from that analysis is 
given in Table~3, together with our results for the UV absorbers, which imply
further inhomogeneities in the outflow in NGC~3783.
To convert the X-ray component solutions for a consistent comparison with the UV
models, we matched models derived with the SED defined in \S 5.1 with the 
model predictions in Paper IV, which used an SED with stronger relative emission 
in the EUV (see \S 5.1).  The ionization parameters were scaled by
matching the peak ions in each ionization component with the predicted
columns in Table 3 of Paper~IV. 
The models computed with the two SEDs were found to give very similar results, 
matching the dominant ions in each component to better than 90\% .

    Table~3 shows the three UV components with relatively high-ionization (1b, 2, and 3) 
share the same ionization parameter as the lowest-ionization region modeled
in the X-ray (hereafter XLI), although they have a smaller (integrated) total
column density.
In Table~4, the predicted column densities from the UV absorbers
for key ions with lines in XLI are listed.
This shows the component~3 and component~1b absorbers may give significant 
contribution to some of the XLI lines.  Indeed, upper bounds on $U$, $N_H$ in 
these components are from the measured \ion{O}{7} column density (\S 5.2).  
Table 4 shows component~3 may also contribute strongly to \ion{Mg}{9} and about a 
third of the \ion{Si}{9} and \ion{Si}{10} measured in the X-ray spectrum.
If $N_H$ is near the upper limit for the less well constrained component~1b absorber, 
it may produce strong \ion{Si}{9} -- \ion{Si}{11}, \ion{Mg}{8}, and \ion{Mg}{9}.
\ion{O}{6} is the only ion with lines detected in both bandpasses.  The UV doublet is found to be 
saturated in all three kinematic components (Figure 4b), giving only a lower limit, and thus
consistent with the measurement from the {\it CXO} spectrum, $N_{O VI} \geq$10$^{17}$~cm$^{-2}$ 
(Paper IV).

     The XLI model predicts a \ion{C}{4} column density that
would produce heavily saturated absorption in the UV doublet in all flux states
observed during our monitoring.
However, as shown above, the \ion{C}{4} equivalent width varied in components 2 and 3, implying 
these features are not saturated, at least in the high flux states when their absorption 
was weaker.  Although \ion{C}{4} is likely saturated in component~1, the dominant X-ray absorption 
coincides kinematically with the lower velocity UV components (Paper I).
Our model for component~3, constrained by the unsaturated high-state \ion{C}{4}, 
fails to reproduce the lowest ionization X-ray species measured in Paper IV, underestimating 
\ion{Si}{7}, \ion{Si}{8}, and \ion{O}{5} by factors of 10 or more.  
Component~2 contributes even less due to its relatively small total column density.
We have done extensive modeling to test if this 
discrepancy between the measured UV and X-ray column densities can be reconciled
for any values of the model parameters.
We find no reasonable choice for an SED that reproduces the measured \ion{Si}{7} and
\ion{O}{5} column densities while maintaining unsaturated absorption in the \ion{C}{4} 
UV doublet, due to the similarity in ionization potentials of these ions, nor
does any combination of multiple components with different physical conditions.
One possible explanation for this discrepancy is that it is due to a geometrical effect.
For example, the very large column density of \ion{C}{4} implied by the XLI model  
may be buried in the variable, unsaturated absorption in components 2 and/or 3.
This would require a low covering factor of the UV emission by XLI given
the relatively shallow absorption observed in \ion{C}{4} components 2 and 3 (see Figure 4a).
Since the X-ray emission region is much more compact than the UV BLR (and likely
the UV continuum source as well), it is reasonable that the same gas would have different 
line-of-sight covering factors in the two bandpasses.
Qualitatively, this is consistent with the inhomogeneous wind model described in detail in the next
section, in which denser, lower-ionization regions occupy smaller and smaller volumes in the global outflow.
The lowest ionization X-ray species could conceivably come from a region that is sufficiently dense
and compact that its presence is not seen against a larger absorber
giving the unsaturated \ion{C}{4}.
Alternatively, it may imply the XLI absorber does not occult the UV absorber at all
and that we are seeing physically distinct absorbers in different lines-of-sight in
the two bandpasses, thus placing very specific requirements on the absorption -- emission
geometry.

  Finally, we compare independent constraints on the distances of the absorbers based
on the UV and X-ray analysis. 
In Paper IV, limits on variability in the X-ray absorption were used to
give lower limits of $R \geq$ 3, 0.6, 0.1~pc for XLI, XMI, and
XHI, respectively (see Table 3).  Similar lower limits were determined from variability 
analysis of {\it XMM-Newton} observations by Behar et al. (2003).
In contrast, a recent study by \citet{kron05} reported variations in the
Fe M-shell unresolved transition array in the {\it CXO} observations, and
they derive an upper limit of $\approx$~6~pc for this absorber.  
Limits can also be derived on the distances for each of the ionization 
components modeled in Paper~IV based on the simple geometrical requirement that 
$\Delta R \leq R$.
If the absorbers are uniform, constant density regions, then 
for a given model solution with parameters $U$ - $N_H$, 
$\Delta R \propto R^2$ through their dependences on density.
This gives upper limits on $R$ (lower limit on $n_H$); results are listed in Table~3.
Filling factors $<$1 would imply more stringent limits on $R$, 
since the absorbers would then occupy a larger region than 
given by $\Delta R$.
The limit on XHI, $R \leq$~4~pc, is somewhat less than
the distance derived for component 1a based on the \ion{C}{3}* density
constraint and UV modeling above; 
it is similar to the estimates for UV components 2 and 3 based on the
assumption of uniform absorbers, with roughly spherical geometries (see discussion in \S 6.1).

\subsection{Global Model of the Outflow in NGC 3783}

\subsubsection{Evidence for an Inhomogeneous Wind}

   With constraints derived on the physical state and geometry of the absorbers,
we now investigate implications for the global model
of the outflow in NGC 3783 and explore constraints for dynamical models.
In Paper IV, the three ionization components modeled for the X-ray absorption
were found to be consistent with being in pressure equilibrium ($P \propto T/U$), 
all lying on stable regions of the nearly vertical part of the 
thermal stability curve (log($T$) vs log($U$/$T$); see Figure 12 in Paper IV).
Based on the modeling above, UV kinematic components 1b, 2, and 3 are 
also at the same pressure.
They occupy the low-temperature base of the region of the thermal stability curve where
a range of temperatures can co-exist in pressure equilibrium, coinciding
with the solution in log($T$), log($U$/$T$) for the low
ionization X-ray gas, XLI, seen in Figure 12 in Paper IV.
Thus, these results are consistent with the general model presented
for NGC 3783 in Paper~IV, and described theoretically in \citet{krol95,krol01},
in which the absorption arises in a multi-phase thermal wind, comprised of embedded 
regions that are inhomogeneous in temperature and density.
In this model, the UV absorbers (and XLI) represent the lowest-ionization, densest material
detectable in the spectrum.

  These results can be compared further with the recent study by \citet{chel05},
which presents a new
dynamical model for the X-ray outflow in NGC 3783 that combines some of
these general ideas with more specific assumptions and calculations. The
main driver of the flow in this model
is thermal gas expansion and the main carrier of the flow is the highest
ionization, hottest component. The model suggests that cooler, lower-ionization
material occupies smaller fractions of the flow where the
size distribution resembles what is known from the ISM.  
An inhomogeneous wind model opens the possibility for
having different covering fractions for different ionization components 
that share the same outflow velocity.
It also eases the problem of the large transverse dimension of the
X-ray gas in component XLI discussed in Paper~IV, since
the low-ionization components can have large column densities yet they 
are composed of small filaments or clouds with relatively small dimensions.
In the framework of this model, \citet{chel05} compute the kinetic 
energy associated with the NGC 3783 outflow to be only
a small fraction of the bolometric luminosity, and the mass-loss rate to be 
comparable to the accretion rate.

    The low-ionization, high-velocity UV absorber component~1a does not fit into
the picture of inhomogeneities co-existing at pressure equilibrium.
It has a gas pressure that is a factor of ten greater than the
other components and thus, if embedded in the more diffuse higher-ionization
gas without an additional confining mechanism, will eventually evaporate.
One possibility is that component~1a is comprised of relatively high density 
material that has recently been swept up from an external mass source
and exposed to the ionizing radiation from the AGN, 
and is destined to expand to come into pressure equilibrium 
with the remainder of the flow.
This absorber was found to have the smallest covering factor, based on the
\ion{Si}{4} doublet absorption, consistent with this being a dense, compact region 
in the flow.
Perhaps it is embedded in, and evaporating into the more diffuse component 1b
gas, which is at pressure equilibrium with the rest of the outflow.
This would explain the dynamical link between these regions implied
by the decrease in radial velocity observed in both absorbers (Paper III).
If component~1a is unconfined, the timescale for it to expand, with its 
density decreasing to that of the 1b region, can be 
approximated based on the estimated size of the absorber
($\Delta R \approx$ 2$\times$10$^{16}$~cm) and the thermal velocity
($v_{th} =$~22 km~s$^{-1}$ at the model temperature
of 1.9$\times$10$^{3}$~K), giving a lifetime of $\sim$ 150 years.
Alternatively, component 1a (and all UV absorbers) could be confined by
magnetic pressure, as proposed in some dynamical models \citep[e.g.][]{emme92,deko95}.
Only a moderate field ($B \approx$ 10$^{-3}$~G) would be required to balance 
the thermal pressure of this absorber; general calculations by \citet{rees87} show fields of
this strength could easily be present at the distance derived for the UV absorbers
in NGC 3783.

    The independent appearance of the UV components
on yearly timescales, without correlation with the observed continuum flux (KC01), 
provides further constraints for physical models of the outflow.
In the framework of the thermal wind model described above, 
it may be a signature of condensations in the outflow that have appeared
in our line-of-sight to the AGN.  However, we note these timescales for 
the appearance of absorption components are quite short compared with the evaporation timescale
derived for component 1a above.
Alternatively, it may be due to motion of the UV absorbers across
our sightline, as discussed in KC01 and Paper III.
In this scenario, dynamical models must account for the implied transverse
component of velocity ($v \geq$ 500 km~s$^{-1}$, KC01; Paper III), 
at the $\sim$ 10 pc distance scale, as well as the radial velocity for the UV absorbers.
Additionally, the decreasing radial velocity observed in component~1 
must be explained, consistently for both subcomponents 1a and 1b.
If this is due to a geometrical effect, in which the absorber is following a
curved path across our line-of-sight, it provides
constraints on its trajectory and kinematics (see Paper III).

  Finally, the combined UV and X-ray results on the outflow geometry and kinematics 
may provide additional constraints.  
If all ionization components are roughly co-located, then the gas giving rise
to the UV absorbers would occupy a much smaller region than the high-ionization material
seen in the X-ray.
If the UV absorption components are single uniform clouds, it may be difficult to account for 
the similarity in outflow velocities for the full range of ionization seen in the two bandpasses;
it would require three small, localized regions (the UV absorbers) span the full kinematic range of the much more
extended global outflow corresponding to the higher ionization gas.
One possibility is that the individual components have low internal filling factors, 
so that the lower-ionization gas we see is distributed over a larger volume, rather than localized 
to the small region implied by $\Delta R$ derived above.
Another potential explanation is that the gas is not co-located (e.g., see discussion in \S 6.2). 
This could be explained, for example, 
by dynamical models that predict the terminal velocities of flow regions will depend on
the launch radius of the outflow, such as the MHD wind models presented in \citet{bott00}; 
in this framework, it is possible that gas with a range of physical conditions and at different locations
would have similar terminal velocities.

\subsubsection{Connection to the Observed Emission-Line Spectrum}

   Here we explore a potential connection between the UV absorbers and the optical -- UV 
line emission.
In a recent study, \citet{cren05} argued the intrinsic UV absorbers 
in Seyfert galaxies may be identified with a high-ionization component of gas in 
the inner narrow line region (NLR), 
based on their kinematics, ionization, inferred distance, and global covering factor.
This is supported by our results for NGC 3783.
From STIS slitless spectra of a sample of Seyfert galaxies, 
\citet{cren05} found 9 of 10 objects have bright, central compact
[\ion{O}{3}] emission-line knots with half-width at half-maximum sizes 
of $\sim$ 5--70~pc and half-width at zero intensity radial velocities 
of 300--1100~km~s$^{-1}$.
The derived distances and observed outflow velocities for the intrinsic 
UV absorbers in NGC 3783 are consistent with this inner NLR component.
A further comparison can be made with the line-emission observed in NGC~3783.
Comparing to the measurements in \citet{evan88},
we find the UV absorbers may contribute significantly to some of the high
ionization UV--optical emission-line fluxes.  
For component~1a, the most constraining line is [\ion{Ne}{5}]~$\lambda$3426,
with a luminosity of $L \approx$10$^{41}$~ergs~s$^{-1}$,
corrected for the NLR reddening determined by \citet{ward84}.
This can be compared to the prediction from component~1a, giving 
$L =$3$\times$10$^{41}$~ergs~s$^{-1}$~$\times C_g$,
where $C_g$ is the global covering factor of the absorber.
This limits $C_g \leq$ 0.3 for gas with similar physical conditions
as component~1a.
This component also predicts a strong contribution to the [\ion{O}{3}]~$\lambda$5007 
emission, and the coronal iron lines ([\ion{Fe}{6}]~$\lambda$5177, [\ion{Fe}{7}]~$\lambda$6085, 
[\ion{Fe}{10}]~$\lambda$6375, and [\ion{Fe}{11}]~$\lambda$7891), reproducing 30 -- 100\% of their 
observed line luminosities for $C_g =$1.
Absorption component~3, with its relatively high-ionization and large column
density, is also predicted to contribute strongly to the higher-ionization coronal iron lines;
e.g., it produces $\sim$3 times the observed [\ion{Fe}{11}]~$\lambda$7891 luminosity
for full global coverage, thereby limiting $C_g \leq$0.3.
It may also contribute to the observed high-ionization UV resonance lines, particularly 
\ion{O}{6}~$\lambda\lambda$1032,1037, implying a similar limit on $C_g$, with only a small
contribution ($\approx$ 5 -- 10\%) coming from resonance scattering.
These constraints on $C_g$ for the absorbers in NGC 3783 are similar to the general constraint  
implied by the detection rate of UV absorption in Seyfert 1s \citep[$\sim$50\%,][]{cren99}.
Finally, we note there is potential evidence for the connection between the UV absorbers and
high-ionization line emission in the profiles observed in NGC 3783.
\citet{ward84} found a strong asymmetry towards blue wavelengths in the forbidden 
coronal iron lines, with emission extending to $\approx-$1500~km~s$^{-1}$, which is not present 
in the low-ionization NLR lines such as [\ion{S}{2}] (see their Figure 6).

\section{Summary}

    We have presented analysis of the intrinsic UV absorption in the 
Seyfert 1 galaxy NGC 3783, based on an intensive monitoring 
campaign with {\it HST}/STIS and {\it FUSE}.
Our 18 STIS observations included both an intensive phase (3 -- 8 day sampling) 
and a long-term phase (several months sampling), observing the UV continuum in
a range of flux states spanning a factor $\sim$~2.5.
In the three kinematic components with strong UV absorption
(components 1, 2, and 3 at radial velocities $v_r \approx -$1350,
$-$550, and $-$725~km~s$^{-1}$, respectively), the equivalent widths 
of the lowest-ionization lines present were observed to vary inversely 
with the UV continuum flux.
This indicates all components are responding to the ionizing flux,
including the short timescales of the intensive observations.

   We isolated the NLR emission using variability in the emission-line 
profiles, in order to explore its effect on the line-of-sight absorption covering factors.
We find a coverage model with an unocculted NLR and 
individual covering factors of the continuum and 
BLR sources derived from the Lyman lines 
predicts saturation of several lines that have substantial residual
fluxes.
These results are consistent with the independent evidence for
saturation in these lines from the lack of observed variability 
and predictions from modeling results.
This coverage model is used to obtain ionic column densities in 
low and high state spectra for photoionization modeling.
Based on differences in derived covering factors and kinematic 
structure, the absorption associated with component~1 appears 
to be comprised of two physical components (1a and 1b).
The decrease in radial velocity observed in lines associated with
both regions implies they are linked dynamically and co-located.

     We were able to resolve the \ion{C}{3}*~$\lambda$1175 absorption complex
observed in component~1a into its individual multiplet lines and measure 
column densities for each metastable level.
Comparing the ratios of the level populations with our calculations,
we determine the electron density for this absorber to be 
$\sim$2.5$\times$10$^4$~cm$^{-3}$.

   We compare the measured ionic column densities with photoionization 
models to constrain the physical conditions in each UV component,
incorporating variability observed between low and high states.
Based on the density from \ion{C}{3}* and the modeling solution, 
component~1a is found to be located at $\sim$ 25 pc.
It is in a relatively low-ionization state (log($U$)$=-$1.6)
compared with the other UV components.
Using the density derived for this absorber, we compute time-dependent
ionic populations and are able to reproduce the
detailed variability observed in this component.
Components 1b, 2, and 3 are found to have similar ionization parameters 
(log($U$)$= -$0.5 -- $-$0.4), with total column densities spanning
a range of a factor $\sim$5.
The ionization parameter for these components is the same as that
modeled for the lowest-ionization X-ray region from the {\it CXO} spectrum, 
but with smaller total column density.
This may indicate different covering factors in the two bandpasses, perhaps
due to the different sizes and line-of-sight geometry of the background emission regions.
Analysis of the observed variability in components 2 and 3
limits their distances from the central ionizing source to be 
$\leq$ 25 and $\leq$ 50 pc, respectively.

    The modeling results for the higher ionization UV components (1b, 2, 3)
imply pressures similar to those of the three X-ray ionization components modeled
in Paper IV.
Thus, the global outflow in NGC 3783 is consistent with an inhomogeneous wind,
with relatively small dense regions giving the UV and lower-ionization X-ray 
absorption, which are embedded in a 
hotter, more tenuous flow.
However, component~1a has a factor of 10 greater pressure than the other absorbers,
thus, if embedded in a larger outflow and unconfined it will eventually 
evaporate, over a timescale of $\sim$150 years.
Alternatively, this absorber could be confined by another mechanism, such as magnetic
pressure.  
The emission-line luminosities predicted from our models of the UV absorbers
indicate they may be contributing substantially to the high-ionization
lines observed in the UV-optical spectrum.
The implied constraints on the global covering factors of the UV absorbers in NGC 3783
are similar to those implied by the detection rate of intrinsic absorption in Seyfert 1
galaxies.
Additionally, the distances derived for the UV absorbers are consistent with the location of 
the inner narrow line region.

   Support for proposal 8606 was provided by NASA through a grant from the Space
Telescope Science Institute, which is operated by the Association of
Universities for Research in Astronomy, Inc., under NASA contract NAS 5-26555.
Some of the data presented in this paper were obtained from the Multimission
Archive at the Space Telescope Science Institute (MAST). Support for MAST for
non-HST data is provided by the NASA Office of Space Science via grant
NAG5-7584 and by other grants and contracts.  We are grateful to Anand Bhatia
for insightful discussions and to Gary Ferland for providing us with an
updated version of Cloudy.

\clearpage

\begin{deluxetable}{lcrl}
\tablecolumns{4}
\footnotesize
\tablecaption{Archival UV Spectra of NGC~3783 \label{tbl-1}}
\tablewidth{0pt}
\tablehead{
\colhead{Instrument/} & \colhead{Coverage} &
\colhead{Resolution} & \colhead{Date} \\
\colhead{Grating} & \colhead{(\AA)} &
\colhead{($\lambda$/$\Delta\lambda$)} & \colhead{(UT)}
}
\startdata
IUE SWP    &1150 -- 1978 & $\sim$250 & 1978 November 26 -- 1992 July 29\tablenotemark{a}\\
FOS G130H & 1150 -- 1605 & $\sim$1200 & 1992 July 27\\
GHRS G160M &1527 -- 1597 & 20,000 & 1993 February 5\\
GHRS G160M &1527 -- 1597 & 20,000 & 1994 January 16\\
GHRS G160M &1527 -- 1597 & 20,000 & 1995 April 11\\
STIS G140L &1150 -- 1715 & $\sim$1000 & 2000 May 17\\
\enddata
\tablenotetext{a}{See the {\it IUE} Merged Log at
http://archive.stsci.edu/iue.}
\end{deluxetable}

\begin{deluxetable}{lllllll}
\tablecaption{Measured and Predicted Ionic Column Densities in the UV Absorbers\tablenotemark{a}
\label{tbl-2}}
\tablewidth{0pt}
\tablehead{
\colhead{Ion}  & \multicolumn{2}{c}{Component~1} & \multicolumn{2}{c}{Component~2}  &
\multicolumn{2}{c}{Component~3}\\
\colhead{} & \colhead{Low} & \colhead{High} &  \colhead{Low} & \colhead{High} &  \colhead{Low} &
\colhead{High}}
\startdata
H I & $\geq$350  &  $\geq$240  &  $\geq$20  &  $\geq$15  & $\geq$80   & $\geq$55  \\
& (670) & (420) & (24) & (12) & (120) & (66)  \\
N V  & $\geq$11 & $\geq$11 & 11 $\pm$3 & 4.9 $\pm$1.5  & $\geq$17  & $\geq$17    \\
& (160) & (160) & (9.5) & (3.5) & (60) & (25)  \\
N III  &  $\geq$8  &  4 & $\leq$0.5 & $\leq$0.5 & $\leq$1.5 & $\leq$1.0  \\
& (34) & (10) & ( - ) & ( - ) & (0.05) & (0.01)  \\
C IV  & $\geq$4.5 & $\geq$4.5 & 2.2 $\pm$0.5 & 0.82 $\pm$0.2 & $\geq$10 & 4.1 $\pm$1.4  \\
& (340) & (190) & (2.7) & (0.71) & (21) & (6.1)  \\
C III & \nodata & \nodata & $\leq$0.5  &  $\leq$0.5 & 0.60 $\pm$0.20 & $\leq$0.3  \\
& (82) & (25) & ( - ) & ( - ) & (0.38) & (0.08)  \\
C III*\tablenotemark{b} & 0.8 ($\pm$0.25) &  0.35 ($\pm$0.3) &  $\leq$0.2  & $\leq$0.2 & $\leq$0.2 & $\leq$0.2 \\
& (0.3) & (0.08) & ( - ) & ( - ) & ( - ) & ( - )  \\
C II & 0.48 $\pm$0.2 & $\leq$0.3 & $\leq$0.25 & $\leq$0.25 & $\leq$0.10 & $\leq$0.10  \\
& (0.59) & (0.12) & ( - ) & ( - ) & ( - ) & ( - )  \\
O VI  & $\geq$14  & $\geq$14 & $\geq$12 & $\geq$12 & $\geq$51 & $\geq$51  \\
& (110) & (51) & (2.2) & (1.0) & (1300) & (640)  \\
Si IV & 2.0 $\pm$0.5 & 0.55 $\pm$0.25 & $\leq$0.1 & $\leq$0.1 & $\leq$0.05  & $\leq$0.05  \\
& (1.2) & (0.23) & ( - ) & ( - ) & ( - ) & ( - )  \\
P V  & 0.13 $\pm$0.04 & 0.09 $\pm$0.06 &  $\leq$0.2 & $\leq$0.2 & $\leq$0.2  & $\leq$0.2 \\
& (0.12) & (0.06) & ( - ) & ( - ) & ( - ) & ( - )  \\
\enddata
\tablenotetext{a}{Column densities listed are in units 10$^{14}$~cm$^{-2}$.
All values measured using the 3-$C$ covering factor model,
except lines associated with component~1a (\ion{Si}{4}, \ion{C}{3}*, \ion{C}{2}, and \ion{P}{5}), see text.
Predicted values from best-fit models are listed below each measured column density.
\nodata denotes no measurement available due to blending with other features, ( - ) listed
for model values $\leq$10$^{12}$~cm$^{-2}$. Lower limits are given for saturated lines.}
\tablenotetext{b}{Population of C$^{+2}$ $^3$P$^o$ $J=$2 metastable level.}
\end{deluxetable}


\begin{deluxetable}{lllllll}
\tablecaption{Constraints on Physical and Geometrical Parameters in UV - X-ray Absorbers\label{tbl-3}}
\tablehead{
\colhead{Component} & \colhead{Log($U$)} & \colhead{Log($N_H$)}  & \colhead{Log($n_H$)} &
\colhead{Log($R$)} & \colhead{Log($\Delta R$)\tablenotemark{a}}  & \colhead{Log($X_T$)}  \\
\colhead{}  & \colhead{}  & \colhead{(cm$^{-2}$)}  & \colhead{(cm$^{-3}$)}  &  \colhead{(cm)}
&  \colhead{(cm)}  &  \colhead{(cm)}
}
\startdata
UV  Component~1a   &   $-$1.6  & 20.6  &  4.40 & 19.94 & 16.2 & \nodata \\
UV  Component~1b  &   $-$0.4  & 20.3 -- 21.4 & 3.20 & 19.94\tablenotemark{b} & 17.1 -- 18.2 &  $\geq$16.0 \\
UV  Component~2   &   $-$0.45  & 20.4 & $\geq$3.3 &  $\leq$19.9  & $\leq$17.0  & $\geq$16.3 \\
UV  Component~3   & $-$0.5  & 21.1 &  $\geq$2.9 &  $\leq$20.2 & $\leq$18.2  & $\geq$16.2\\
X-ray LI\tablenotemark{c} & $-$0.5  & 21.9  & 0.6 -- 4.7   &  19.0 -- 21.3   & $\geq$17.2   & \nodata \\
X-ray MI\tablenotemark{c} & 0.8  & 22.0 &  2.0 -- 5.0     &  18.3 -- 19.9 &   $\geq$17.0     & \nodata \\
X-ray HI\tablenotemark{c} & 1.3  & 22.3 &  3.2 -- 5.4     &  17.7 -- 19.1   & $\geq$16.9  \nodata    & \nodata \\
\enddata
\tablenotetext{a}{Radial depths assuming uniform, constant density absorbers.}
\tablenotetext{b}{Assumed co-located with component~1a based on decrease in radial velocities
of lines in both components, see text.}
\tablenotetext{c}{Results for the three ionization X-ray components from modeling the {\it CXO} spectrum 
in Paper IV; $U$ values converted for models with the SED used in the present study.
Upper (lower) limits on $n_H$ ($R$, $\Delta R$) based on limits on variability from analysis
in Paper IV; lower (upper) limits based on $\Delta R \leq R$.}
\end{deluxetable}


\begin{deluxetable}{lllllll}
\tablecaption{Predicted Ionic Column Densities for Lines in X-ray Spectrum\tablenotemark{a} \label{tbl-4}}
\tablewidth{0pt}
\tablehead{
\colhead{Ion}  & \colhead{Comp~1a} & \colhead{Comp~1b}\tablenotemark{b} &  \colhead{Comp~2}  & \colhead{Comp~3} & \colhead{Total UV\tablenotemark{c}} &  \colhead{CXO\tablenotemark{d}} \\
}
\startdata
O V  & 17.0  & 15.1 -- 16.3 & 15.4 & 16.2  &  17.1  & 17.1 ($\pm$0.40) \\
O VI  & 16.8  & 15.9 -- 17.2 &  16.2 & 17.0  & 17.3 -- 17.5  & $>$17.0 \\
O VII   & 16.4  & 16.8 -- 18.1 & 17.1   &  17.8    & 17.9 -- 18.3  & 18.04 ($\pm$0.20)  \\
O VIII  &  14.5   &  16.4 -- 17.6 &  16.7  &  17.2    & 17.4 -- 17.8  &   18.63 ($\pm$0.25)   \\
Si VII  &  15.7   & 13.4 -- 14.7 &  13.7  &  14.7   & 15.8  &  16.26 ($\pm$0.20)  \\
Si VIII  &  15.4   &  14.5 -- 15.7 & 14.8   &   15.6    & 15.9 -- 16.1  &   16.66 ($\pm$0.08)  \\
Si IX  &  14.6   &  15.1 -- 16.3 & 15.4   &  16.1   & 16.2 -- 16.5  &   16.66 ($\pm$0.06)   \\
Si X  &  13.5   &  15.3 -- 16.5 & 15.6   &  16.1   &  16.3 -- 16.7  &  16.83 ($\pm$0.05)   \\
Si XI  &  12.1   &  15.1 -- 16.3 & 15.4   &  15.8   &  16.0 -- 16.5 &   16.29 ($\pm$0.06)   \\
Mg VIII & 15.1 & 14.9 -- 16.1  &   15.2   &   16.0   & 16.1 -- 16.4  & 16.58 ($\pm$0.05) \\
Mg IX  & 14.1 & 15.3 -- 16.4  &  15.6   &   16.3   & 16.4 -- 16.7  &   16.39 ($\pm$0.06) \\
\enddata
\tablenotetext{a}{Component~1 -- 3 models are intermediate states between
the low and high states used in modeling each UV component; all values are log(N$_{ion}$).}
\tablenotetext{b}{Range in ionic column densities corresponding to range in $N_H$ for component 1b.}
\tablenotetext{c}{Sum of model predictions from all UV components.}
\tablenotetext{d}{Column densities measured in the {\it CXO} spectrum, from Paper IV.}
\end{deluxetable}


\clearpage
\begin{figure}
\centering
\includegraphics[angle=90,width=13cm]{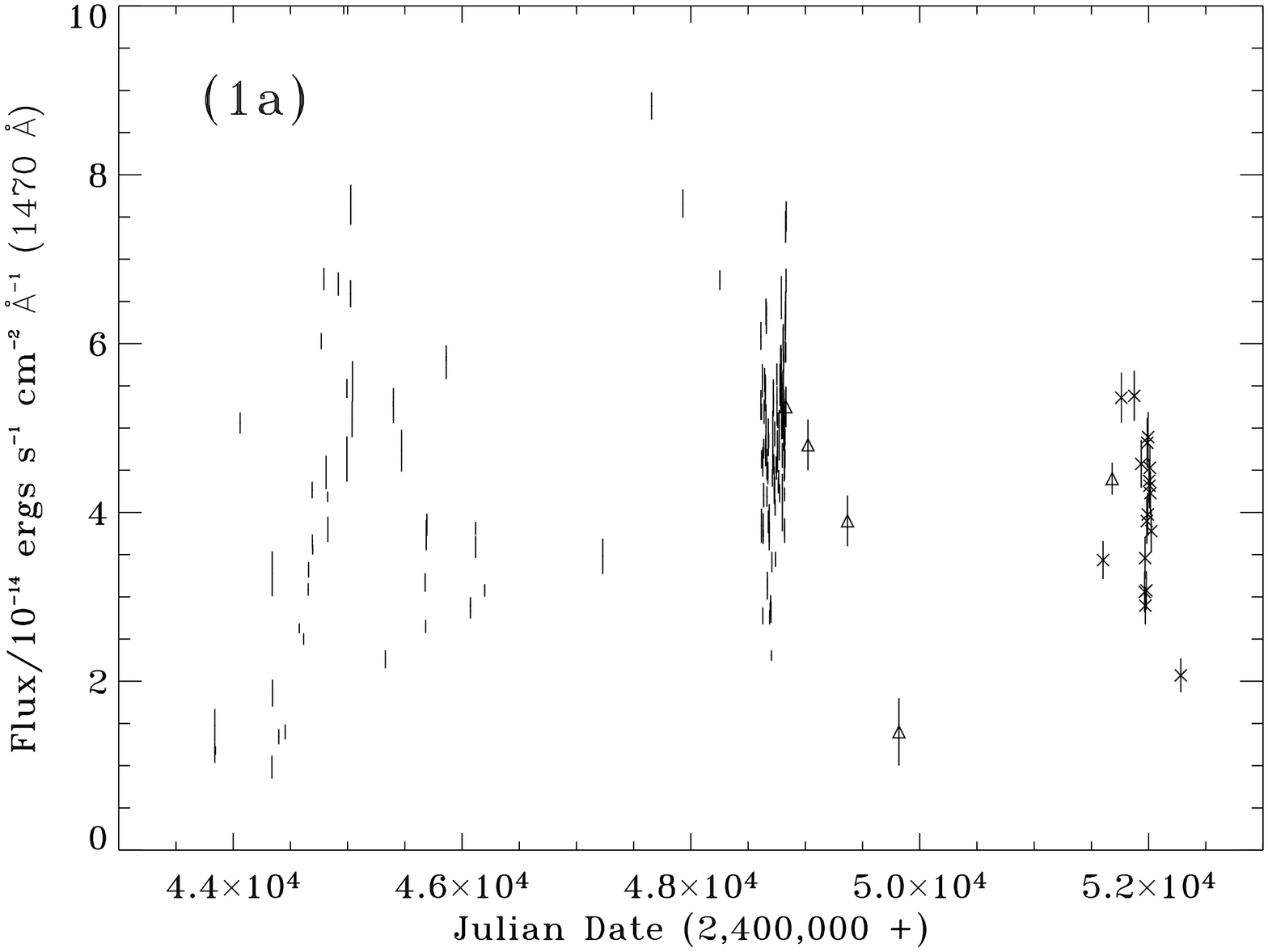}
\vspace*{0.15 in}
\includegraphics[angle=90,width=13cm]{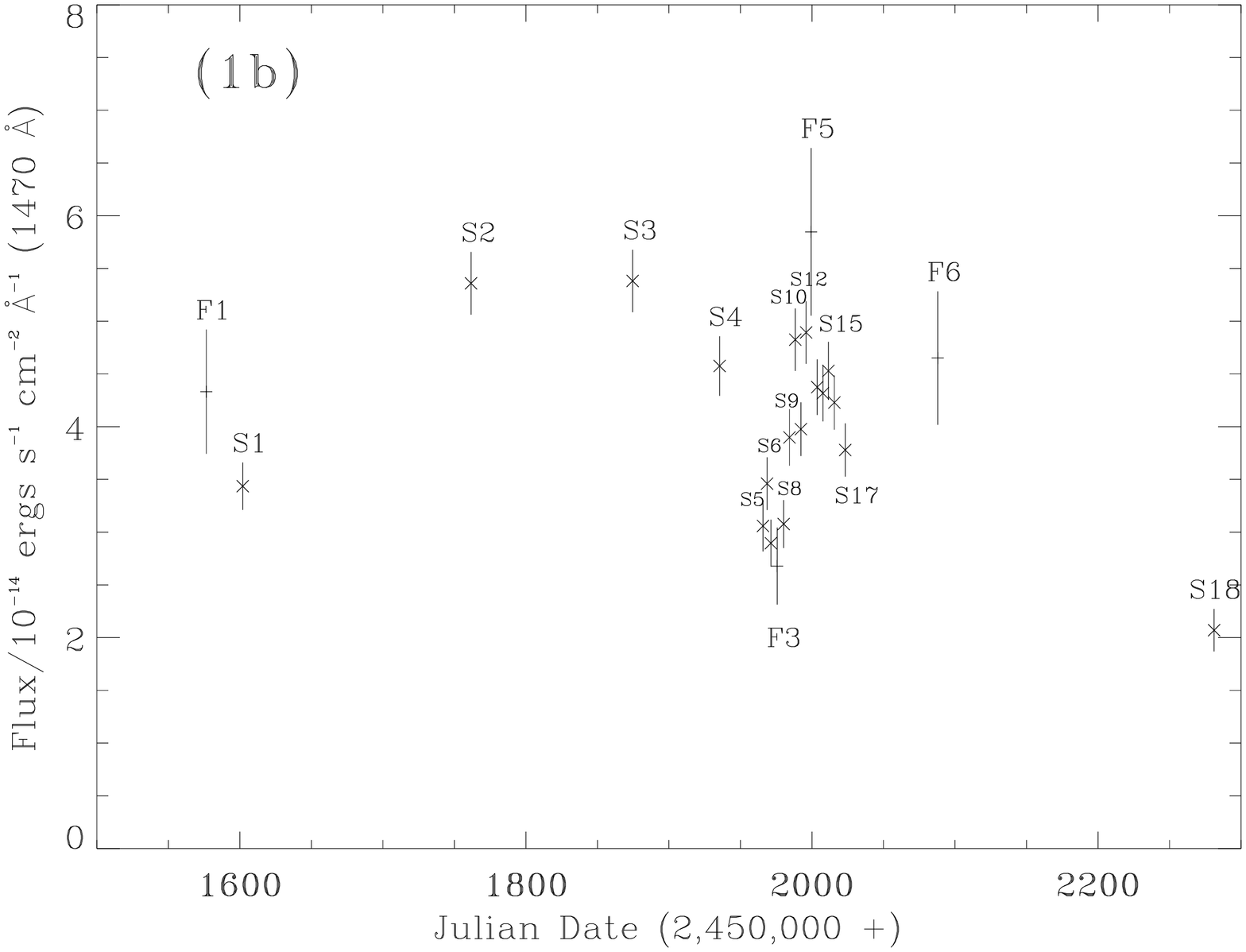}
\vspace*{0.15 in}
\caption{(1a) Continuum fluxes measured at 1470 \AA\ spanning $\sim$22 years of UV observations.
Shown are observations from {\it IUE}, GHRS and FOS ($\triangle$), and STIS ($\times$) with error bars.  
The intensive {\it IUE} (JD~$\approx$~2,448,500) and STIS (JD~$\approx$~2,452,000) observations show 
variations of a factor of $\sim$1.5~--~2.5 are typical on weekly/monthly timescales.  The STIS observations 
span a range of moderately-low -- moderately-high flux states. 
(1b) Continuum fluxes for observations presented in this study.
STIS observations ($\times$) are labeled chronologically S1--S18; fluxes extrapolated
from the {\it FUSE} observations are plotted with ($+$) for the four {\it FUSE} observations that
are non-simultaneous with STIS observations (F1, F3, F5, F6).
The error bars represent estimated 1$\sigma$ uncertainties. \label{fig1}}
\end{figure}

\clearpage
\begin{figure}
\vspace*{1 in}
\includegraphics[width=8.cm]{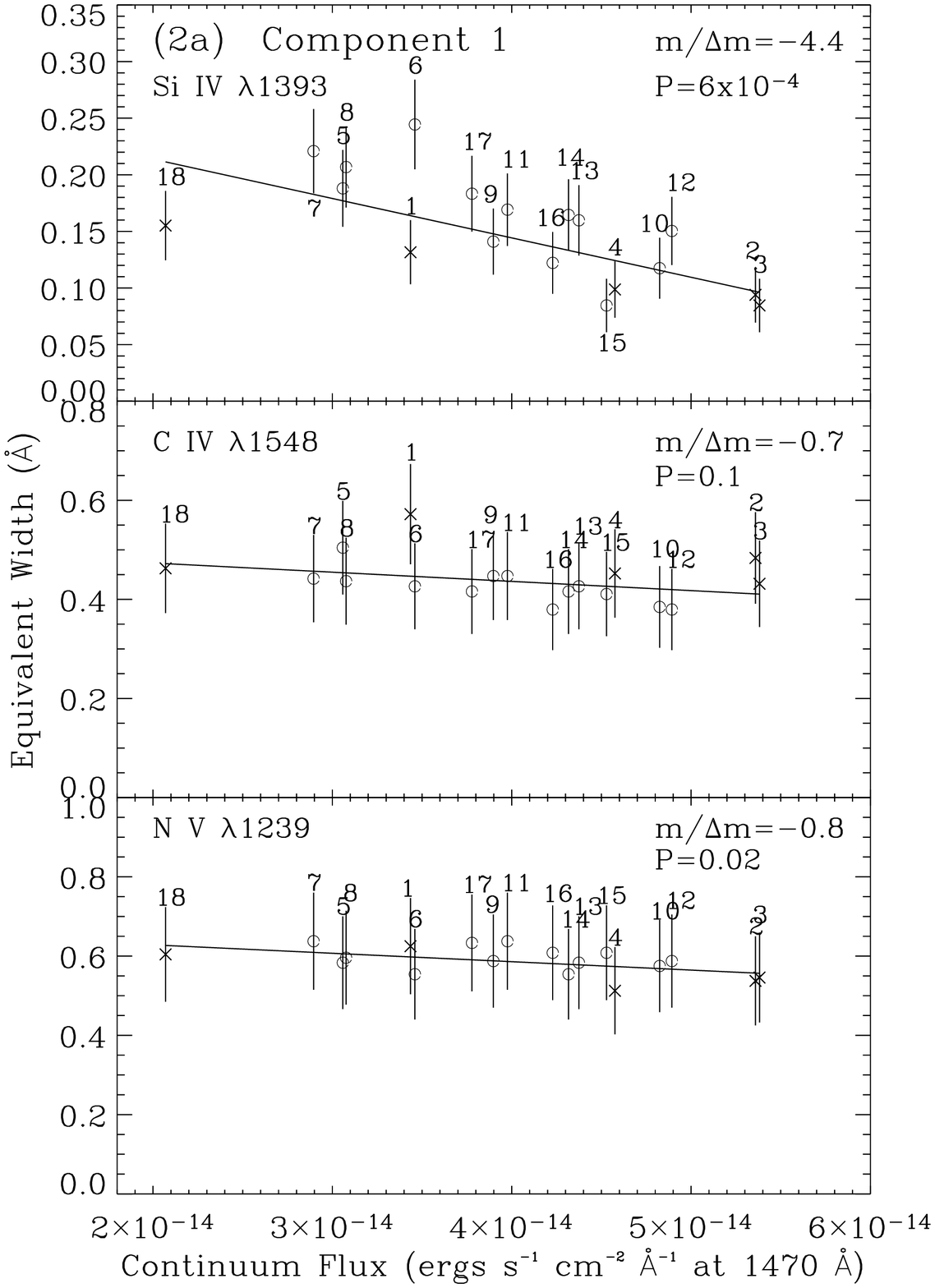}

\vspace*{-4.81 in}
\hspace*{3.5 in}
\includegraphics[width=8.cm]{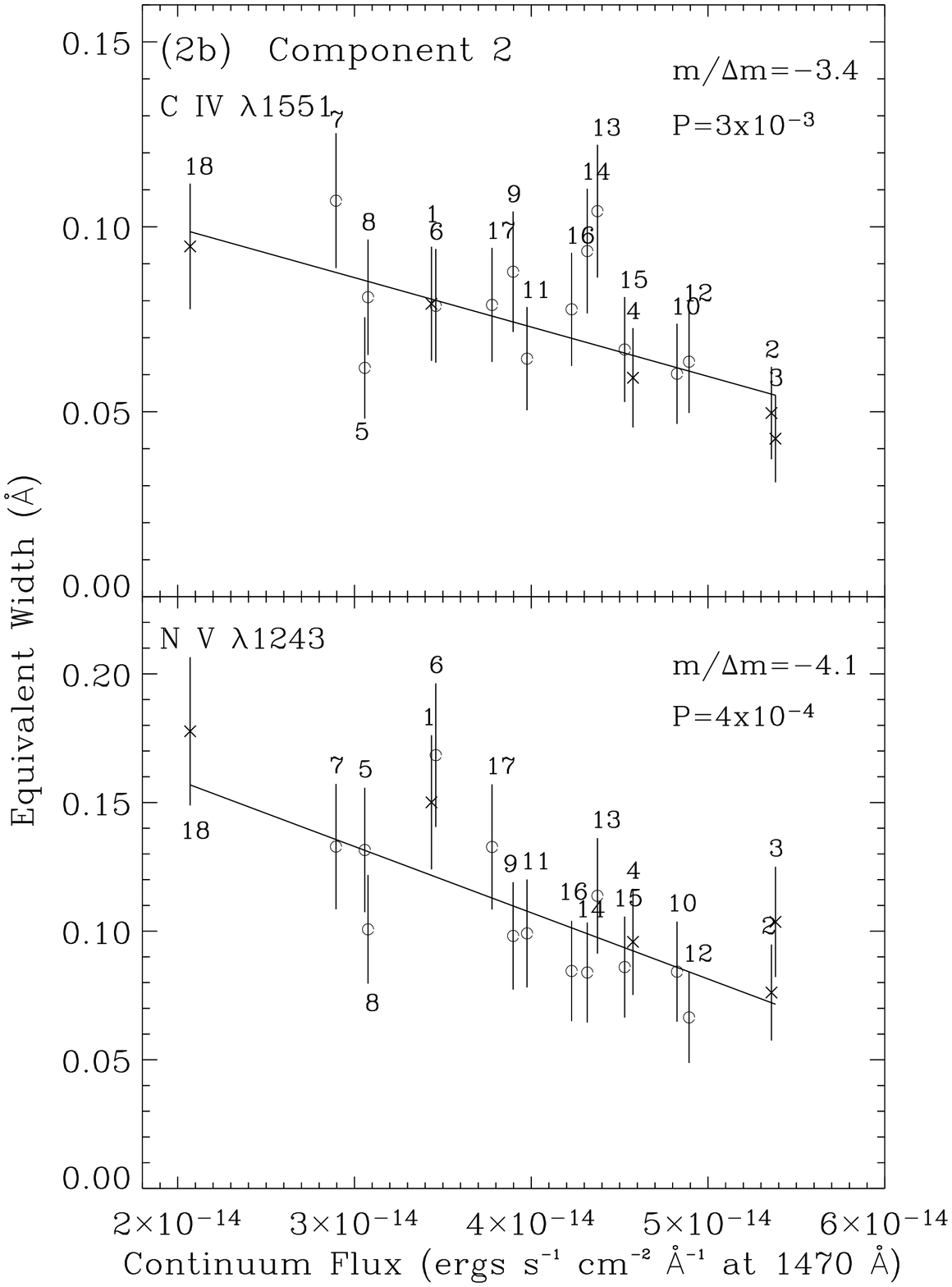}
\vspace*{0.65 in}
\caption{Intrinsic absorption equivalent widths as a function of UV continuum flux from
all STIS observations.
(2a) In component 1, \ion{Si}{4} exhibits a general trend of decreasing
absorption strength with increasing continuum flux, while \ion{C}{4} and \ion{N}{5} show no strong 
variability. 
(2b) \ion{C}{4} and \ion{N}{5} in component 2 are both variable,
showing an inverse correlation with continuum flux.
(2c) The \ion{C}{4} absorption in component 3 shows 
the same inverse correlation with continuum flux, whereas \ion{N}{5} does not vary significantly.
Observations during the intensive
monitoring phase are plotted with ($\circ$) to distinguish them from the long-term 
monitoring ($\times$), and all observations are labeled chronologically for comparison
with Figure 1b.
Results of least squares fits to all epochs are plotted for each line.
The ratio of the computed slope to the uncertainty in slope ($m$/$\Delta m$) is
quoted to characterize the correlation between equivalent width and continuum flux.
Also given are probabilities of the null hypothesis ($P$) from the Spearman rank
correlation test. \label{fig2}}
\end{figure}

\addtocounter{figure}{-1}
\clearpage
\begin{figure}
\centering
\vspace*{1 in}
\includegraphics[width=8.cm]{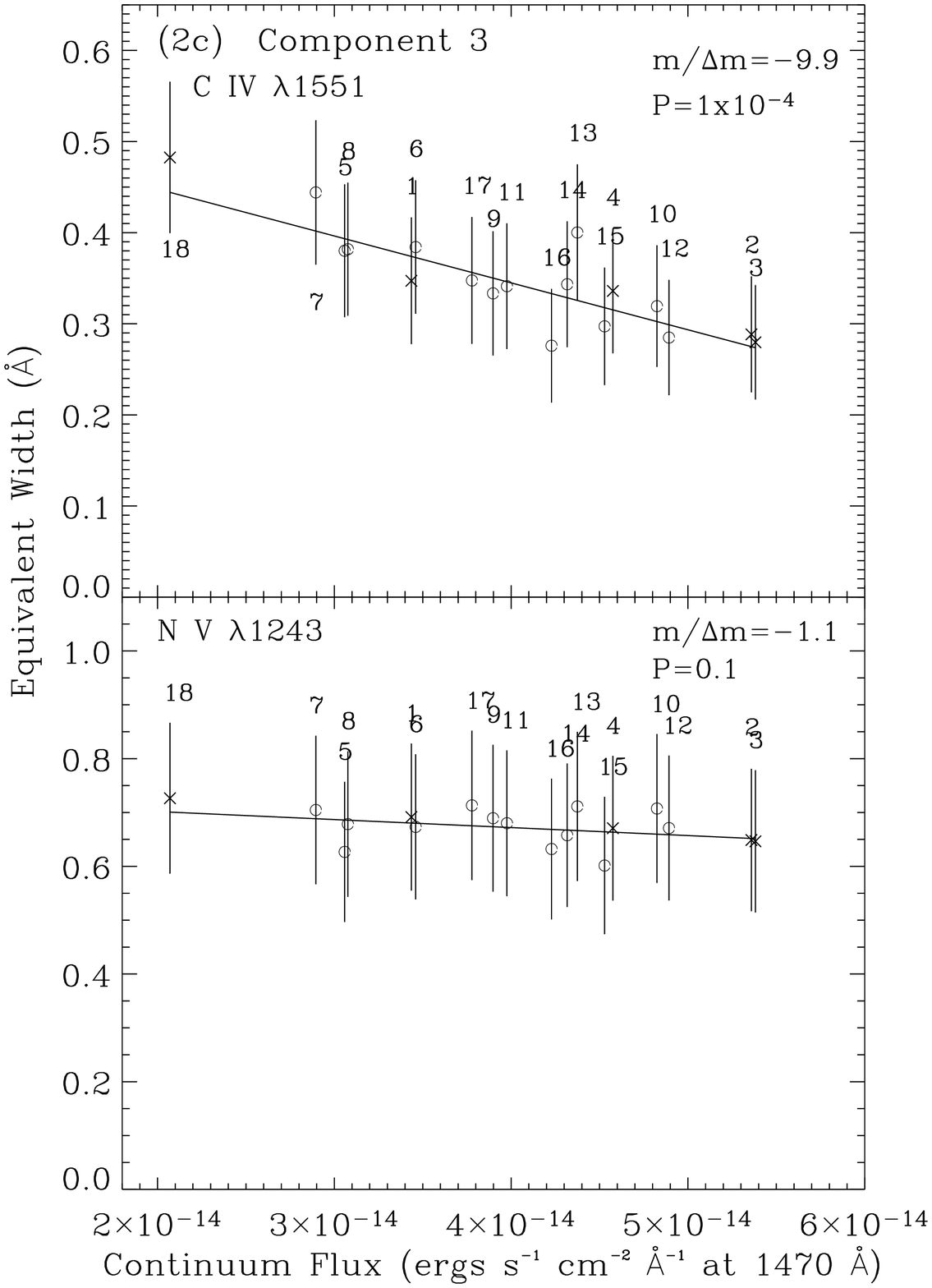}
\vspace*{0.65 in}
\caption{(2c) \label{fig2c}}
\end{figure}


\clearpage
\begin{figure}
\centering
\includegraphics[angle=90,width=13.cm]{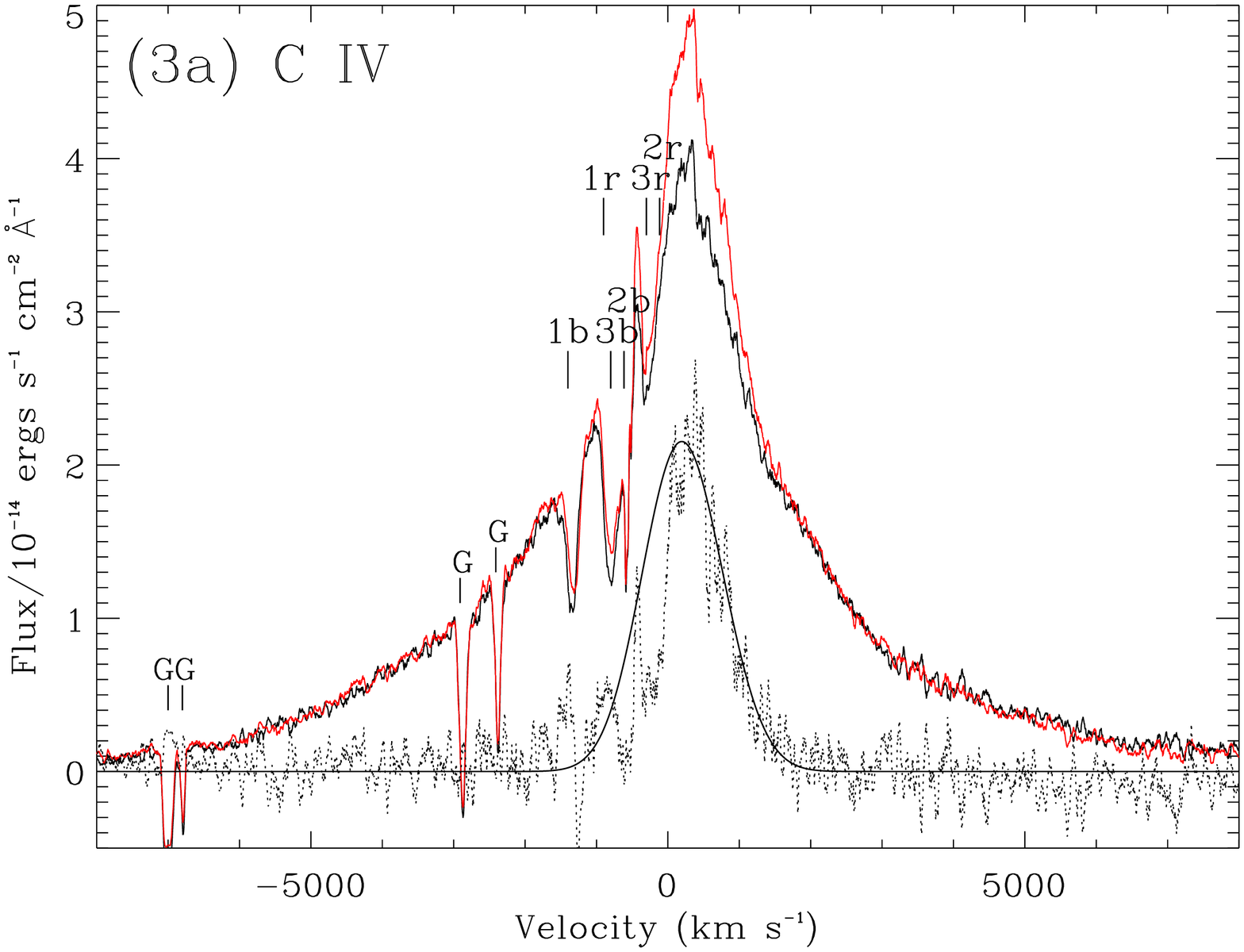}
\caption{Separation of variable and non-variable emission-line components.
(3a)  The high-state (solid black line) and low-state emission-line profiles (grey) 
are shown for \ion{C}{4} after subtraction of the continuum flux.
The low-state profile has been scaled by a factor of 1.4 to match the BLR flux in the high
velocity wings. At lower velocities the profiles diverge, interpreted as the signature of 
a non-varying NLR component.
The solution to the NLR profile is plotted with a dotted line, along with a Gaussian fit 
(smooth black line), as described in the text. The intrinsic absorption components 
for the blue (b) and red (r) \ion{C}{4} doublet members are labeled; Galacitic lines
are denoted with a "G''.
(3b) Similar analysis is shown for \ion{N}{5}, Ly$\alpha$, and \ion{Si}{4}.
The residual NLR profile for each ion is overlaid with a scaled template of the Gaussian fit to the 
\ion{C}{4} NLR profile for comparison.
A detector artifact (art.) and the O IV] emission-line multiplet are identified on the Si IV 
spectrum. \label{fig3}}

\centering
\includegraphics[width=7.cm]{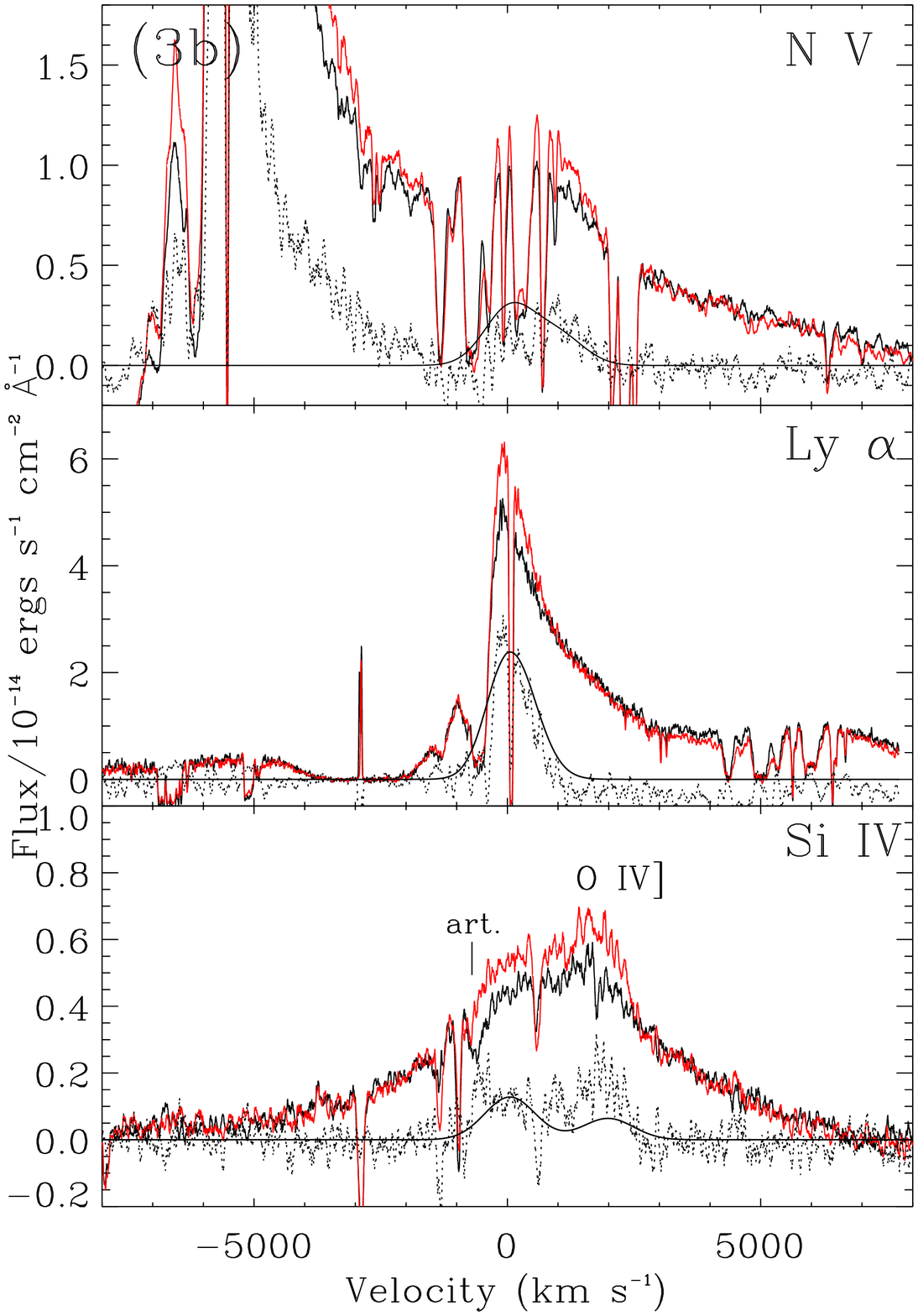}
\vspace*{0.35 in}
\end{figure}

\clearpage
\begin{figure}
\vspace*{1 in}
\includegraphics[width=8.cm]{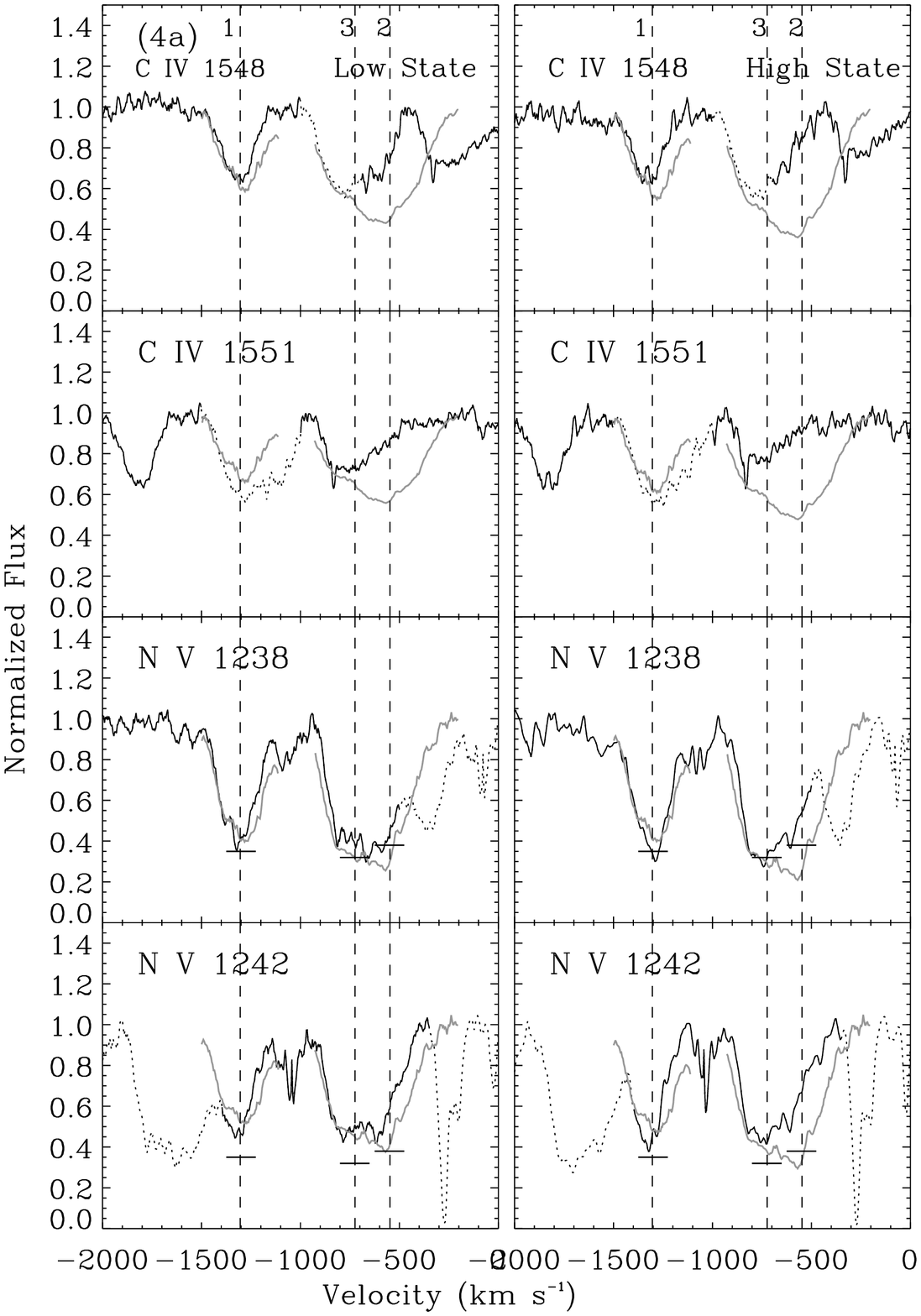}

\vspace*{-4.81 in}
\hspace*{3.5 in}
\includegraphics[width=8.cm]{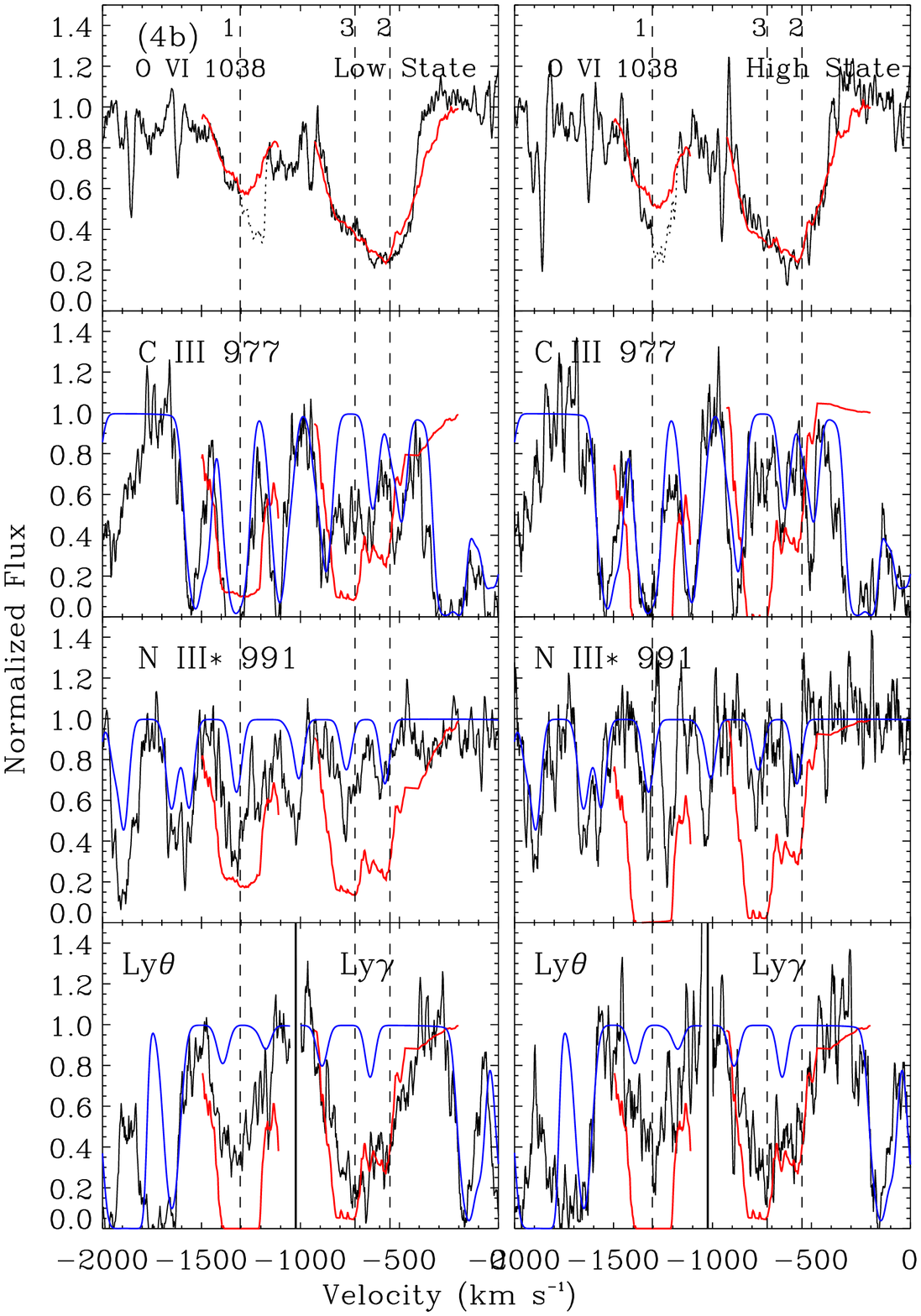}
\vspace*{0.65 in}

\caption{Normalized absorption profiles in mean low (left panels) and high
(right) flux states.  Unocculted flux levels (1 $-C$) from a model 
with an unocculted NLR, and continuum and BLR covering
factor profiles derived from the Lyman lines are shown (thick grey lines in 4a and 4c; red in 4b). 
Spectral regions 
contaminated with other absorption or detector artifacts are plotted with dotted lines.  
A model of the strong Galactic H$_2$ absorption in the {\it FUSE} spectrum is shown (blue line).
Radial velocities of the three strong kinematic components are marked with dashed vertical 
lines. For doublets having both lines free of contamination with other absorption, the
doublet solution to the covering factor is shown with a horizontal bar.  For enhanced S/N,
the C II spectrum shown is the average of the ground state (C II~$\lambda$1335) and fine-structure
(C II*~$\lambda$1336) lines. \label{fig4}}
\end{figure}


\addtocounter{figure}{-1}
\clearpage
\begin{figure}
\centering
\vspace*{1 in}
\includegraphics[width=8.cm]{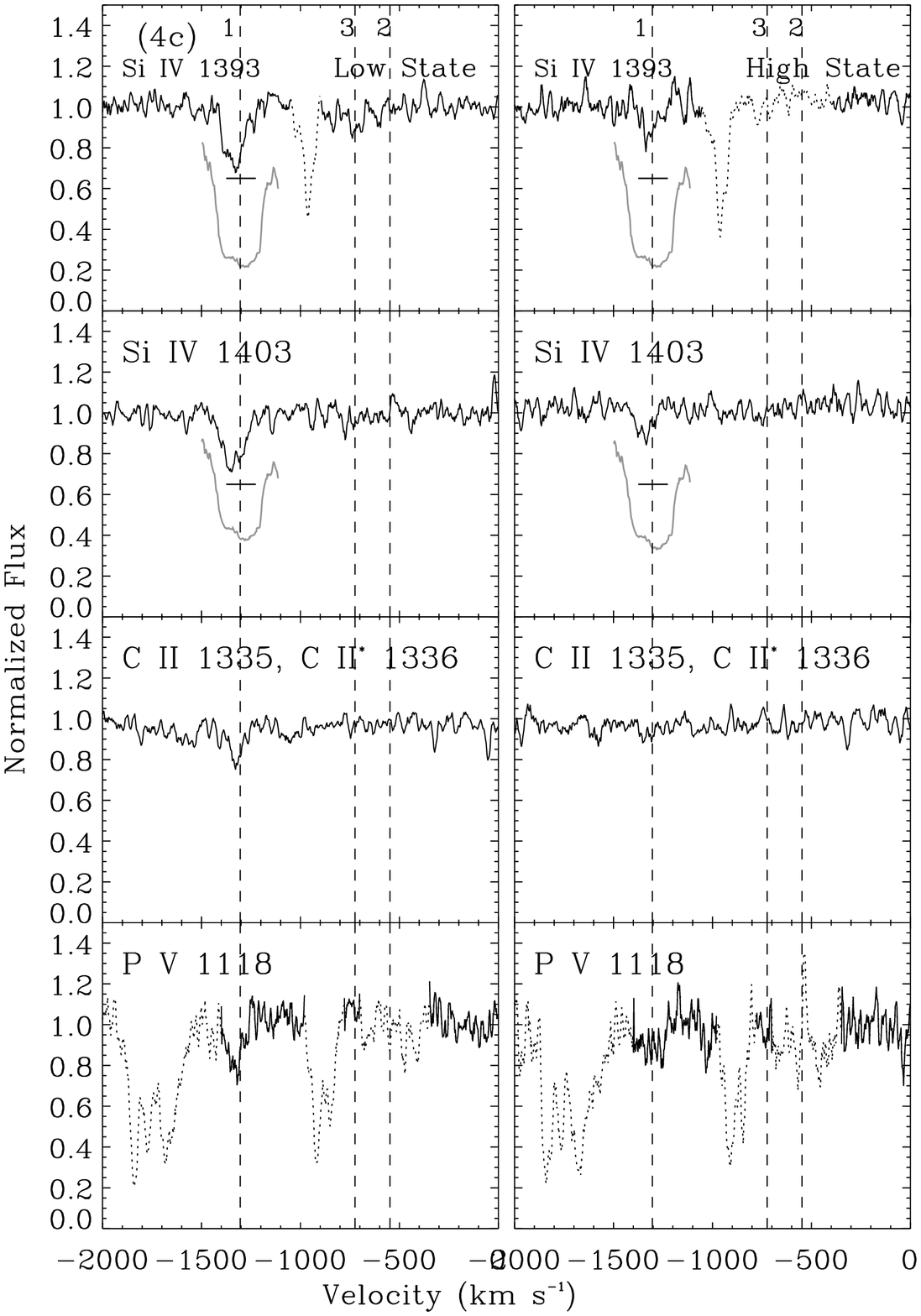}
\vspace*{0.65 in}
\caption{(4c) \label{fig4c}}
\end{figure}


\clearpage
\begin{figure}
\vspace*{0.5 in}
\centering
\includegraphics[width=11.cm]{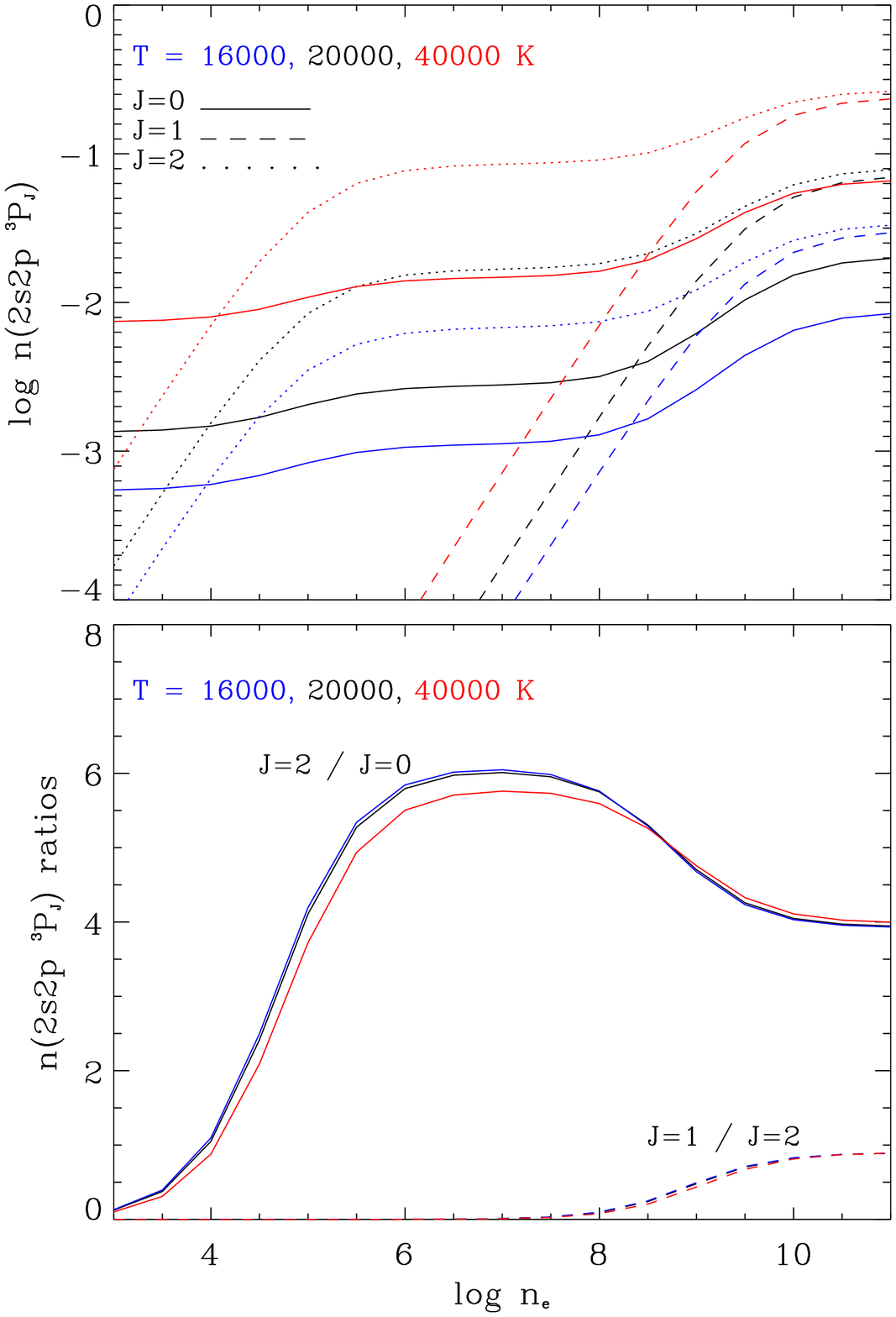}
\vspace*{0.35 in}
\caption{\ion{C}{3} metastable level populations. (Top) The computed
populations for the metastable $^3P_J$ levels of the 2s2p term
of C$^{+2}$ are plotted as a function of electron density for
temperatures of $T =$ 16000 (blue), 20000 (black), and 40000 K (red).
(Bottom) Population ratios for the $J =$ 2:0 (solid lines)
and $J =$1:2 levels (dashed lines) are plotted as a function of
density for the same temperatures.
\label{fig5}}
\end{figure}

\clearpage
\begin{figure}
\vspace*{0.5 in}
\centering
\includegraphics[angle=90,width=16.cm]{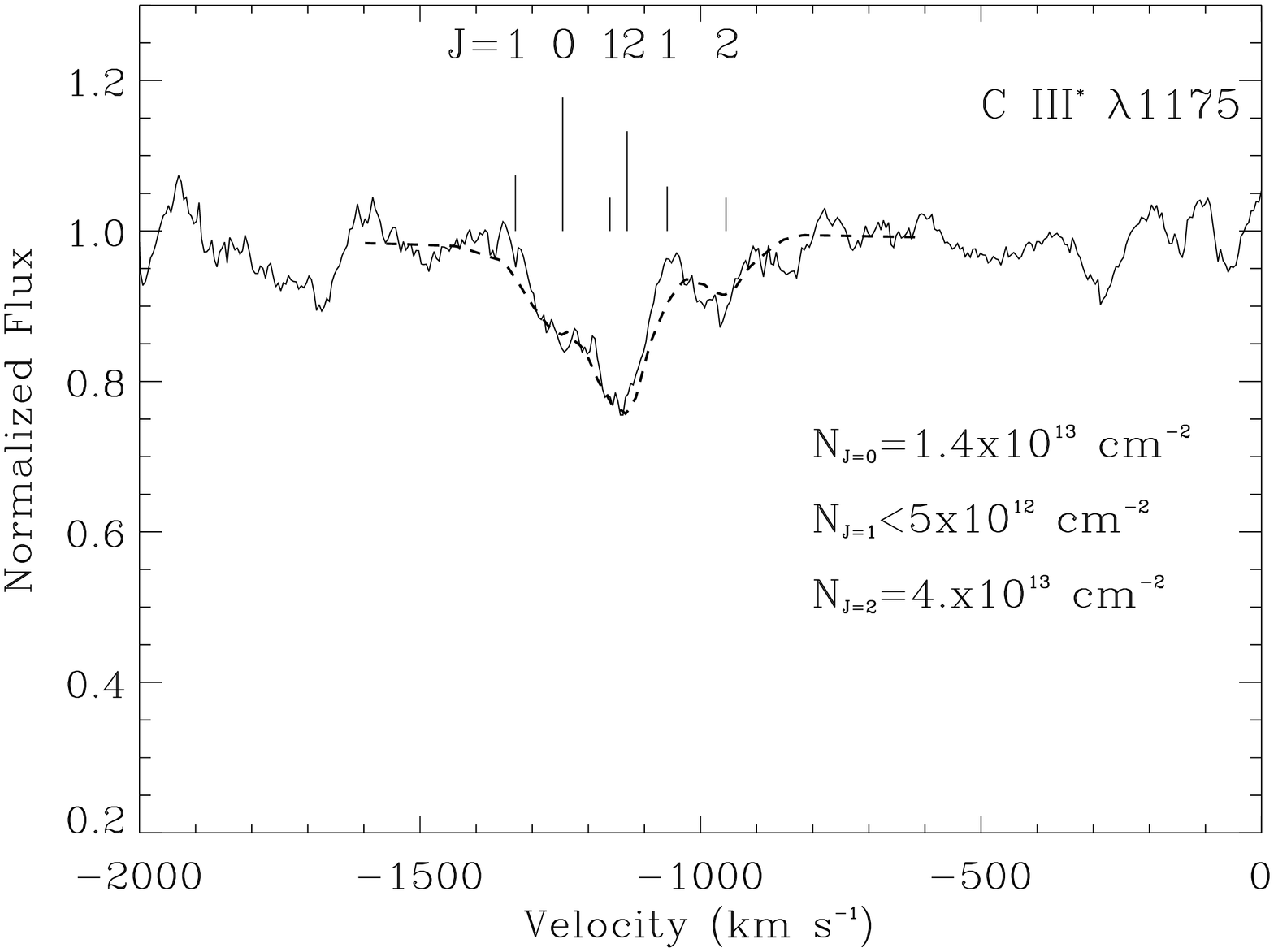}
\vspace*{0.35 in}
\caption{Absorption spectrum for \ion{C}{3}*~$\lambda$1175
multiplet.  The normalized mean STIS spectrum shows
the individual multiplet lines are unblended allowing measurements
of column densities for individual $^3P_J$ levels in the 2s2p term.  
The spectrum is plotted in the velocity frame of the multiplet line 
with shortest wavelength.  The radial velocity of each transition in the 
multiplet corresponding to the centroid of the \ion{Si}{4} feature in the averaged
STIS spectrum is marked with a tickmark above the spectrum, with relative oscillator
strength denoted by line length, and the lower level of each transition
is identified by the $J$ value.
A fit to the absorption complex is plotted with a dashed line, with best-fit column densities 
for each level printed below the spectrum. 
\label{fig6}}
\end{figure}


\clearpage
\begin{figure}
\centering
\includegraphics[angle=90,width=13cm]{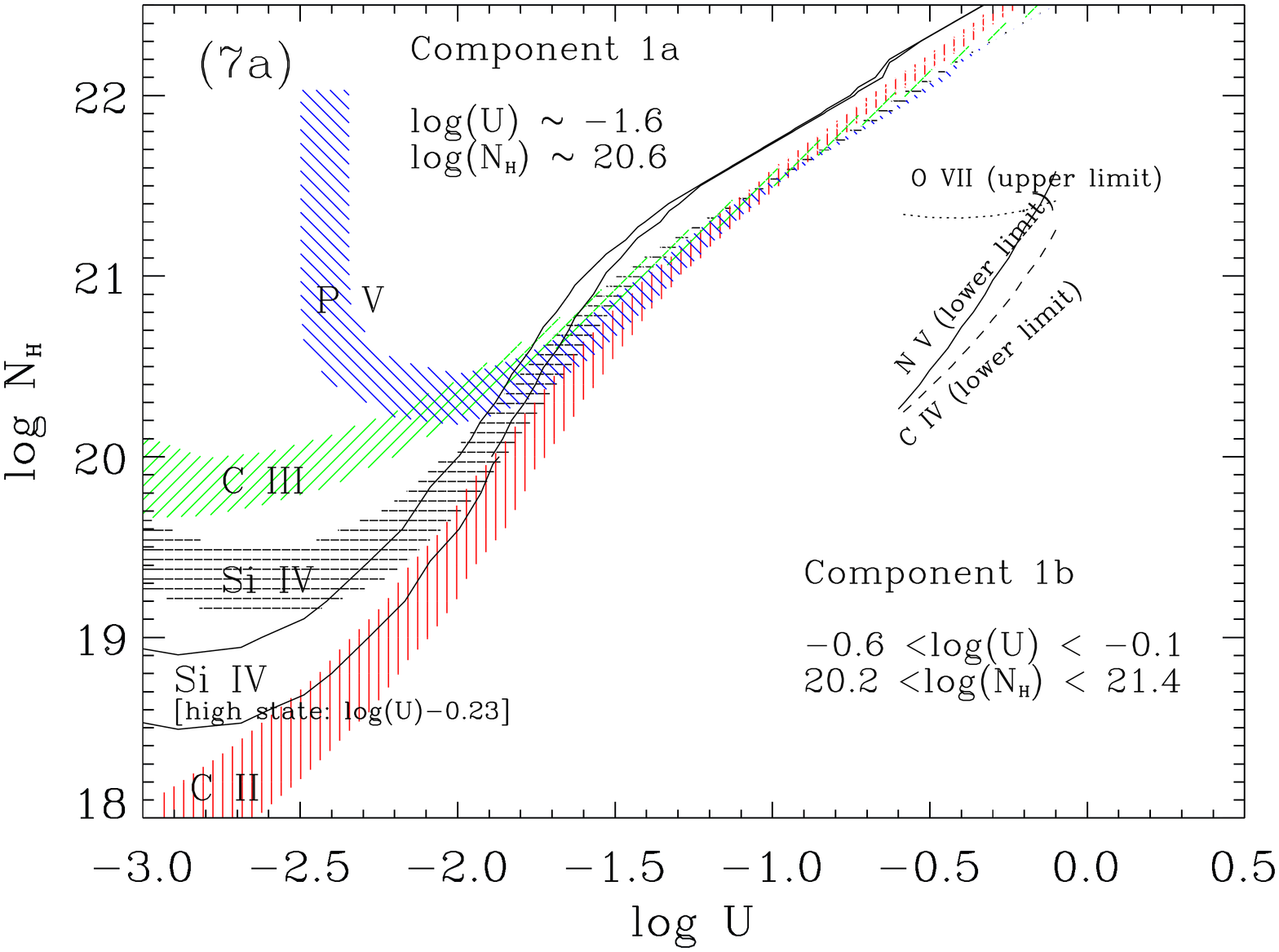}
\vspace*{0.15 in}
\includegraphics[angle=90,width=13cm]{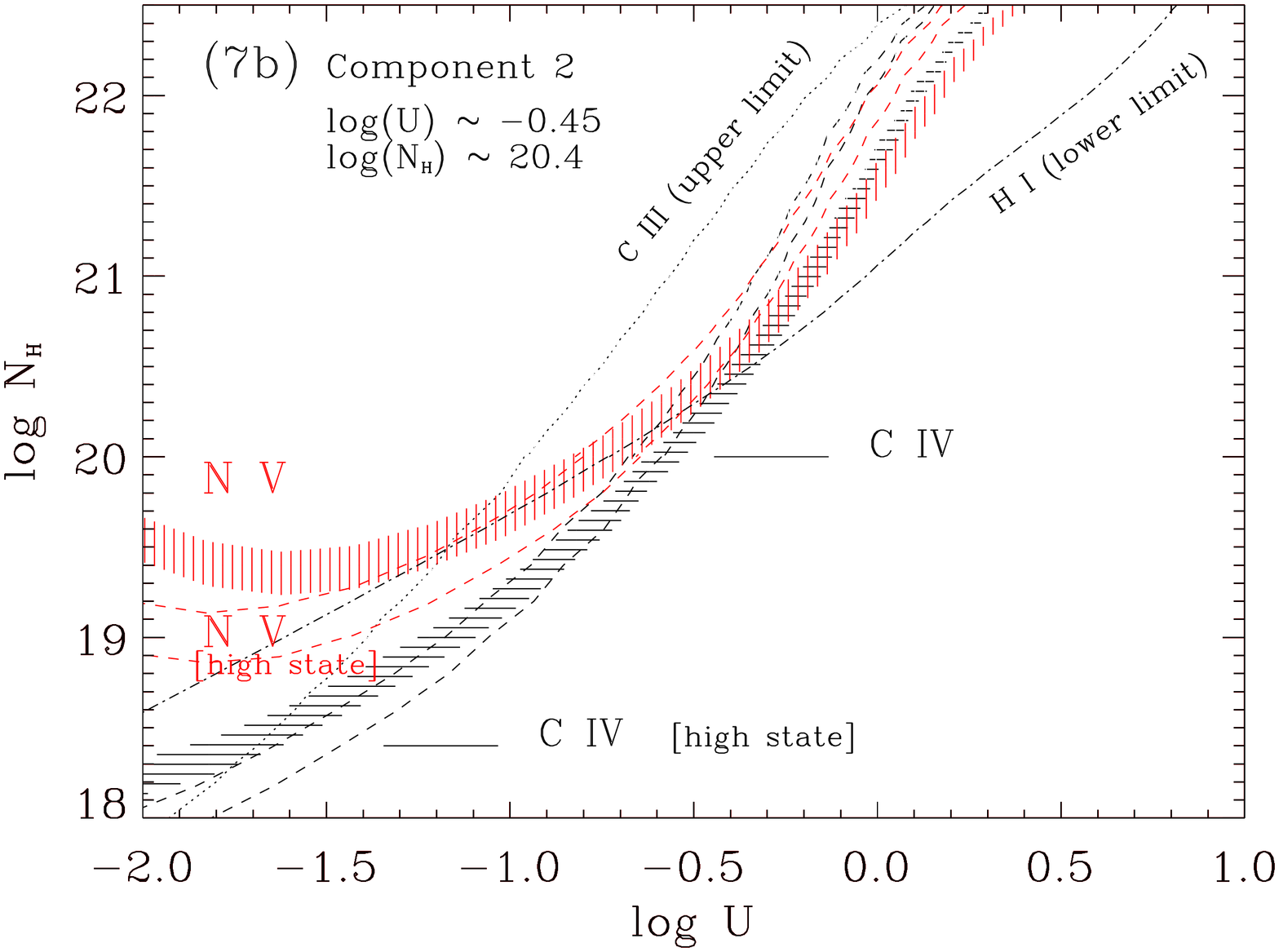}
\vspace*{0.15 in}
\caption{Photoionization modeling results for each intrinsic absorption component.
(7a) For component 1a, solutions in $U$, $N_H$ corresponding to the measured column 
densities in the low-state are shown as contours for each ion.  
The widths span the range of solutions based on estimated uncertainties.
Regions of overlap give models that simultaneously fit those ions.
The high-state solution to \ion{Si}{4} is included (unfilled lines) by shifting the solution
along the x-axis by log($U$)=$-$0.23, corresponding to the continuum flux ratio of the high and low 
states. This allows a direct comparison of solutions in different flux states.  
Model solutions for the measured limits (plotted with single lines) 
for component 1b are shown on the right. 
Solution for components 2 and 3 are shown in Figure 7b and 7c.
\label{fig7}}
\end{figure}


\addtocounter{figure}{-1}

\clearpage
\begin{figure}
\vspace*{1 in}
\centering
\includegraphics[angle=90,width=13cm]{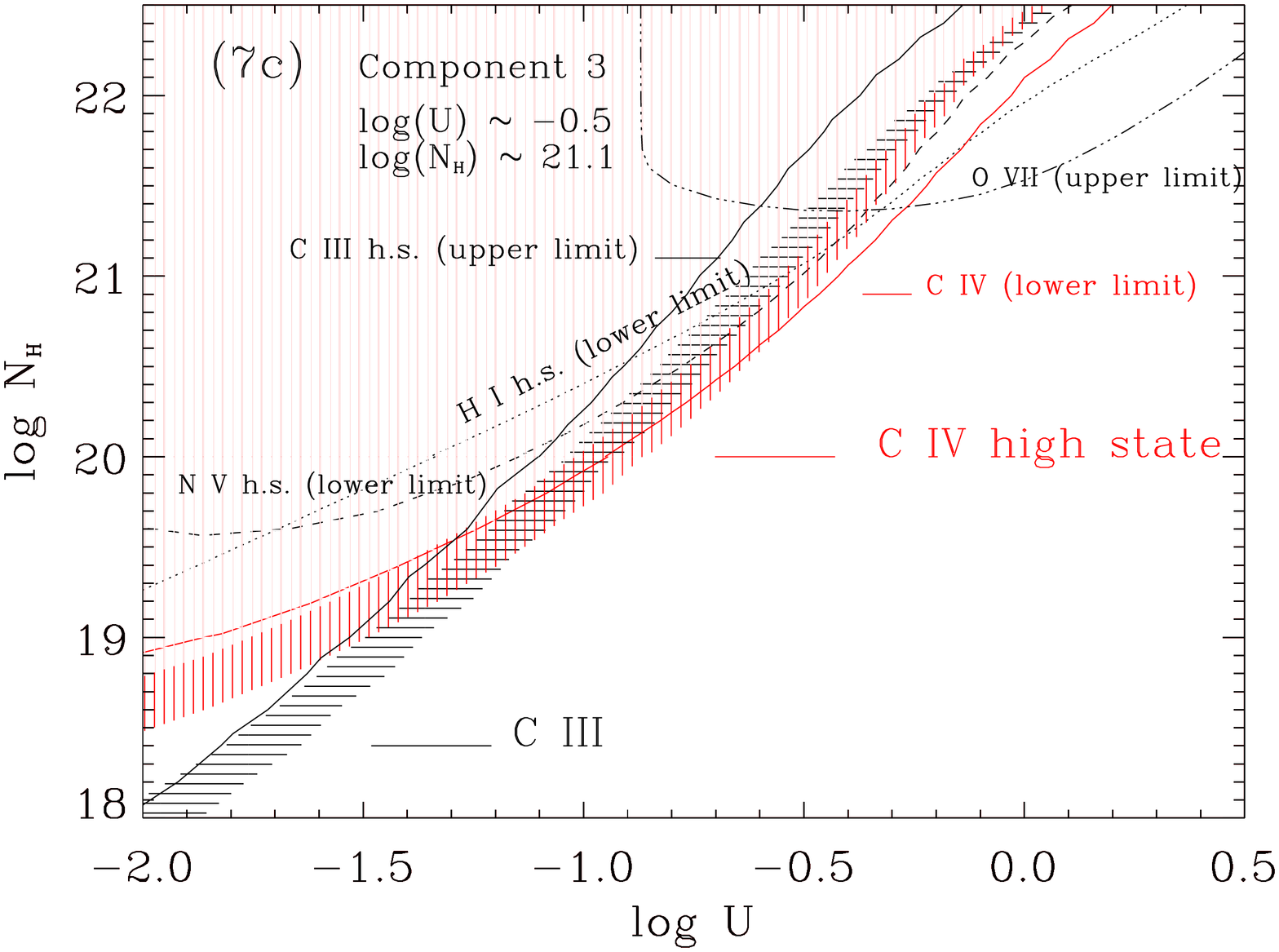}
\vspace*{0.25 in}
\caption{(7c) \label{fig7c}}
\end{figure}

\clearpage
\begin{figure}
\vspace*{0.5 in}
\centering
\includegraphics[angle=90,width=16.cm]{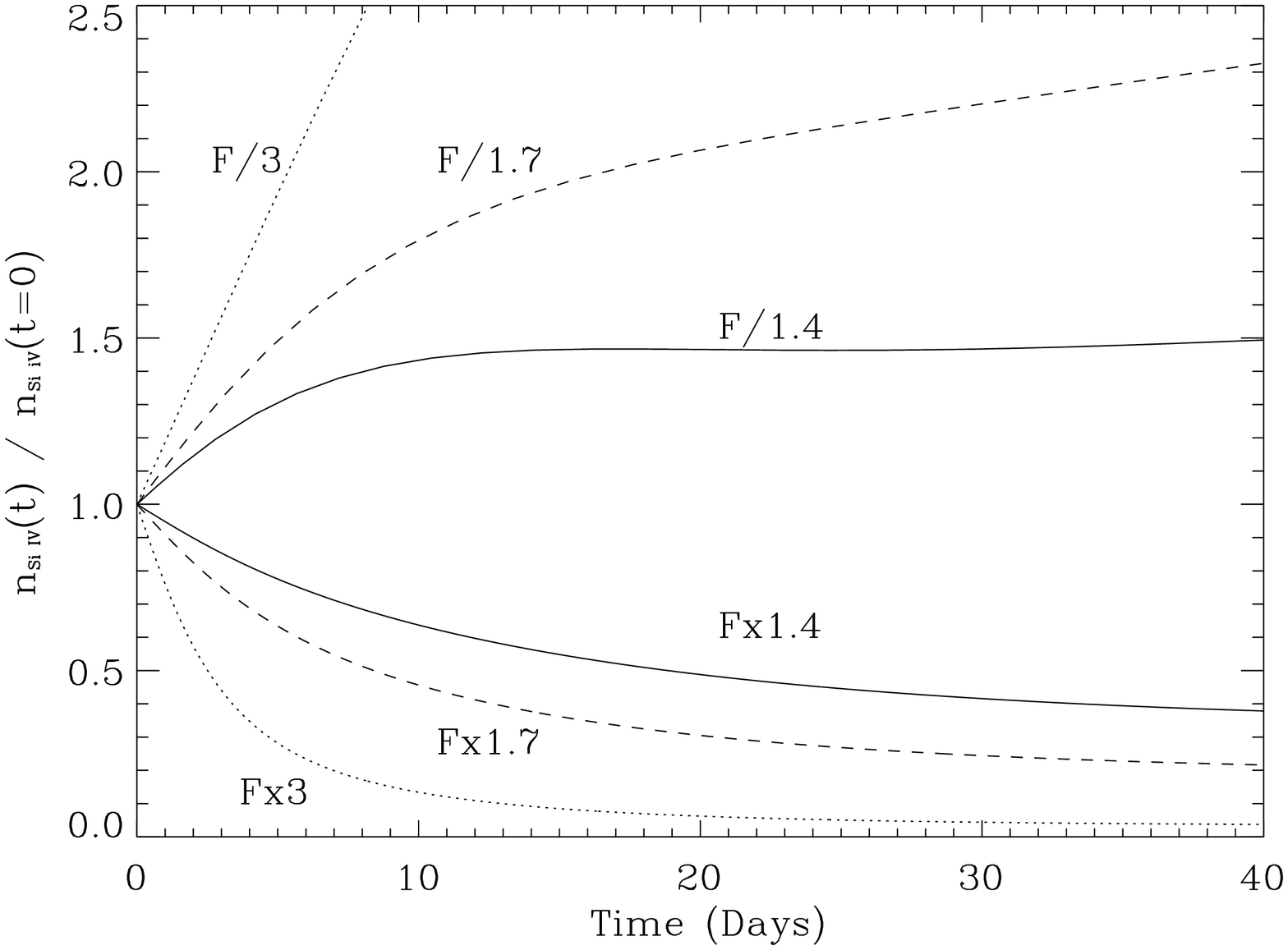}
\vspace*{0.15 in}
\caption{Time-dependent population computed for \ion{Si}{4} in component 1 for various
amplitudes of flux variations.  Results for increases and decreases in ionizing flux by
factors 1.4, 1.7, and 3 are shown.  Flux changes were assumed instantaneous at t=0.
Initial conditions for the calculations are from the best-fit Cloudy model for this
absorber, with the density derived from the C III*~$\lambda$1175 feature.
\label{fig8}}
\end{figure}

\clearpage
\begin{figure}
\vspace*{0.5 in}
\centering
\includegraphics[width=11.cm]{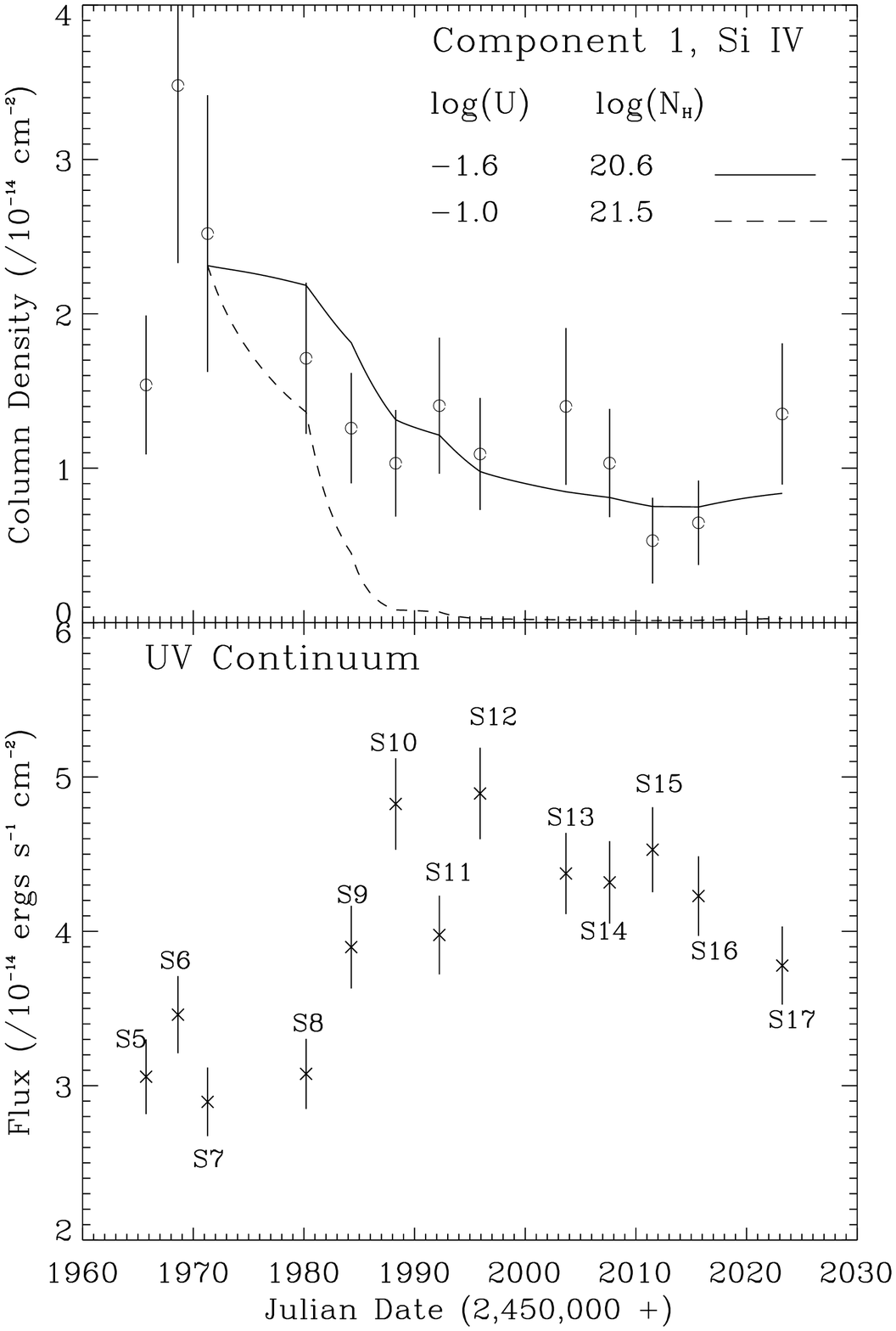}
\vspace*{0.55 in}
\caption{Computed variations in the component 1a Si IV column density during the intensive 
monitoring observations. (Top) Results are shown for
two initial low-state models: the best-fit solution based on simultaneously matching
the low and high-state measurements (solid line)
and a higher-ionization solution from Figure 7a that matched only the low-state column
densities (dashed line). 
The time-dependent populations were calculated between each epoch assuming step-function
variations in ionizing flux equal to the observed UV continuum variability.
Measured column densities are plotted with circles and error bars.  
The overall timescales and magnitudes of variations are matched well by the low-ionization
model, but the variability is severely overestimated by the high-ionization model.
(Bottom) The UV continuum light curve is shown for comparison. \label{fig9}}
\end{figure}

\clearpage
\begin{figure}
\vspace*{0.5 in}
\centering
\includegraphics[angle=90,width=16.cm]{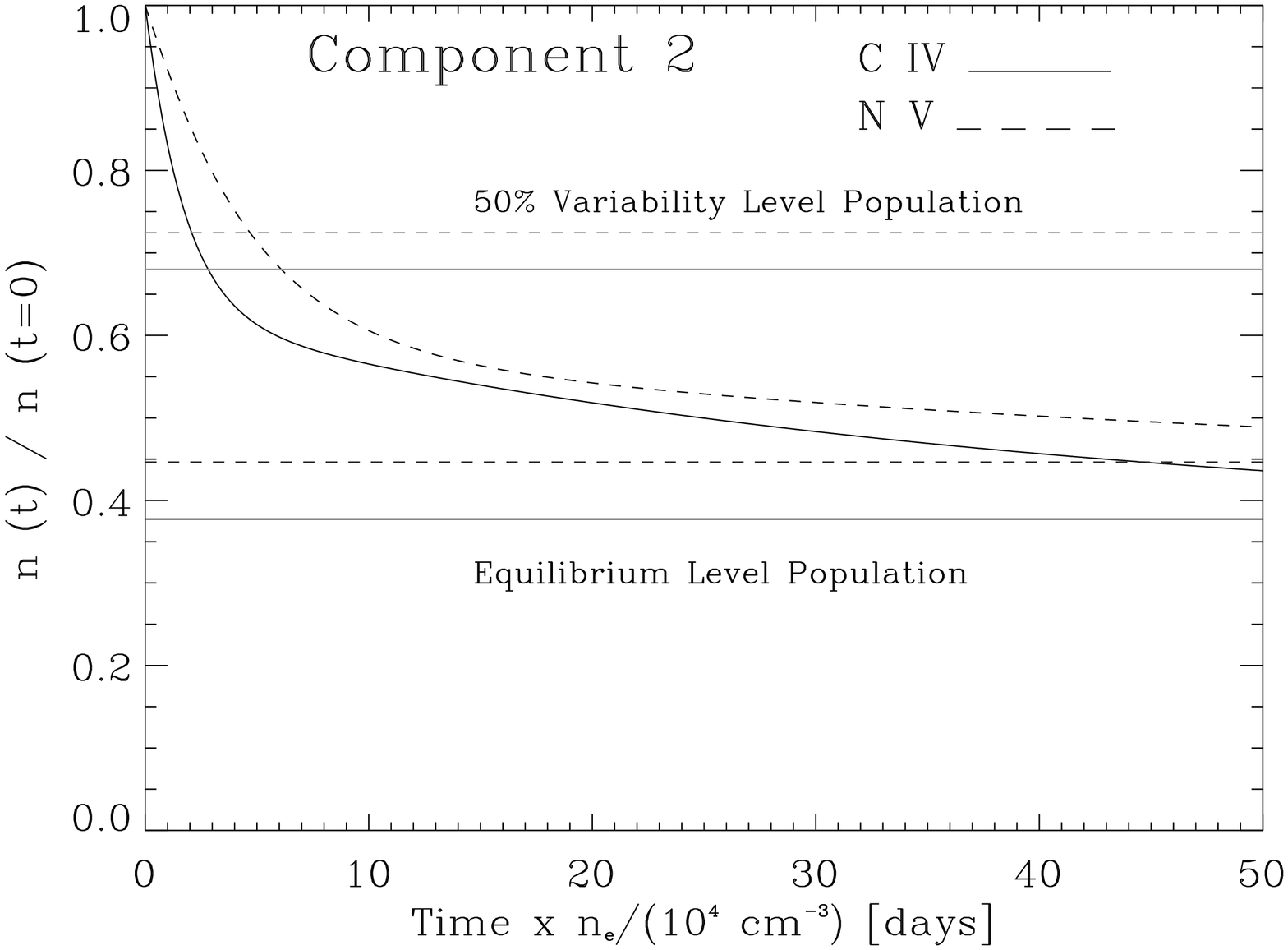}
\vspace*{0.15 in}
\caption{Time-dependent \ion{C}{4} and \ion{N}{5} populations computed for component 2.  
Initial populations are from the best-fit low-state model (see text), with 
a step function increase $\times$1.7 in ionizing flux at $t=$0.
Populations for \ion{C}{4} (solid curve) and \ion{N}{5} (dashed curve) are given relative to initial values,
plotted as a function of the product of time (days) and electron density (10$^4$~cm$^{-3}$).
Lower horizontal lines denote the equilibrium values for the high-state models. 
Upper horizontal lines mark the 50\% variability levels, which are used to
determine lower limits on the density based on the time separation between low
and high states. \label{fig10}}
\end{figure}


\clearpage
\begin{thebibliography}{}
\bibitem[Arav et al.(2001)]{arav01}Arav, N., et al. 2001, ApJ, 561, 118
\bibitem[Barlow \& Sargent (1997)]{barl97}Barlow, T. A., \& Sargent, W. L. W. 1997, \aj, 113, 136
\bibitem[Behar et al.(2003)]{beha03}Behar, E., et al. 2003, ApJ, 598, 232
\bibitem[Berrington et al.(1985)]{berr85}Berrington, K. A., Burke, P. C., Dufton, P. L., \& Kingston, A. E.  1985, Atom. Data Nucl. Data Tables, 33, 195
\bibitem[Berrington et al.(1989)]{berr89}Berrington, K. A., Burke, V. M., Burke, P. C., \& Scalia, S. 1989, J. Phys. B, 22, 665
\bibitem[Bhatia \& Kastner (1993)]{bhat93}Bhatia, A. K. \& Kastner, S. O. 1993, ApJ, 408, 744
\bibitem[Bottorff et al.(2000)]{bott00}Bottorff, M. C., Korista, K. T., \& Shlosman, I. 2000, ApJ, 537, 134
\bibitem[Bromage et al.(1985)]{brom85}Bromage, G. E., et al.\ 1985, \mnras, 215, 1
\bibitem[Cardelli et al.(1989)]{card89}Cardelli, J. A., Clayton, G. C., \& Mathis, J. S. 1989, \apj, 345, 245
\bibitem[Chelouche \& Netzer (2005)]{chel05}Chelouche, D. \& Netzer, H. 2005, ApJ, in press
\bibitem[Clavel et al.(1991)]{clav91}Clavel, J., et al. 1991, ApJ, 366, 64
\bibitem[Crenshaw \& Kraemer (2005)]{cren05}Crenshaw, D. M., \& Kraemer, S. B. 2005, ApJ, in press
\bibitem[Crenshaw et al.(2003)]{cren03}Crenshaw, D. M., Kraemer, S. B., \& George, I. M. 2003, AARA, 41, 117
\bibitem[Crenshaw et al.(1999)]{cren99}Crenshaw, D. M., Kraemer, S. B., Boggess, A., Maran, S. P., Mushotzky, R. F., \& Wu, C.-C. 1999, \apj, 516, 750
\bibitem[de Kool \& Begelman (1995)]{deko95}de Kool, M., \& Begelman, M. C. 1995, ApJ, 455, 448
\bibitem[de Vaucouleurs et al.(1991)]{deva91}de Vaucouleurs, G., de Vaucouleurs, A., Corwin, H. G., Buta, R. J., Paturel, G., \& Fouque, P.  1991 Third Reference Catalogue of Bright Galaxies (Springer-Verlag: New York)
\bibitem[Dietrich \& Wagner (1998)]{diet98}Dietrich, M. \& Wagner, S. J. 1998, A\&A, 338, 405
\bibitem[Emmering et al.(1992)]{emme92}Emmering, E. T., Blandford, R. D., \& Shlosman, I. 1992, ApJ, 385, 460
\bibitem[Espey et al. (1998)]{espe98}Espey, B. R., Kriss, G. A., Krolik, J. H., Zheng, W., Tsvetanov, Z., \& Davidsen, A. F. 1998, \apj, 500, L13
\bibitem[Evans (1988)]{evan88}Evans, I. N. 1988, ApJS, 67, 373
\bibitem[Ferland et al.(1998)]{ferl98}Ferland, G. J., Korista, K. T., Verner, D. A., Ferguson, J. W., Kingdon, J. B., \& Verner, E. M. 1998, \pasp, 110, 761
\bibitem[Gabel et al.(2003a)]{gabe03a}Gabel, J. R., et al. 2003a, ApJ, 583, 178 (Paper II)
\bibitem[Gabel et al.(2003b)]{gabe03b}Gabel, J. R., et al. 2003b, ApJ, 595, 120 (Paper III)
\bibitem[Gabel et al.(2005)]{gabe05}Gabel, J. R., et al. 2005, ApJ, 623, 85
\bibitem[Ganguly et al.(1999)]{gang99}Ganguly, R., Eracleous, M., Charlton, J. C., \& Churchill, C. W. 1999, \aj, 117, 2594
\bibitem[George et al.(1998)]{geor98}George, I. M., Turner, T. J., Netzer, H., Nandra, K., Mushotzky, R. F., \& Yaqoob, T. 1998, \apjs, 114, 73
\bibitem[Grevesse \& Anders (1989)]{grev89}Grevesse, N. \& Anders, E. 1989 in Cosmic Abundances of Matter, ed. C. J. Waddington (New York: AIP), 1
\bibitem[Hamann et al.(1997)]{hama97}Hamann, F., Barlow, T. A., Junkkarinen, V., \& Burbidge, E. M. 1997, \apj, 478, 80
\bibitem[Kaspi et al.(2001)]{kasp01}Kaspi, S., et al.\ 2001, \apj, 554, 216
\bibitem[Kaspi et al.(2002)]{kasp02}Kaspi, S., et al.\ 2002, \apj, 574, 643 (Paper I)
\bibitem[Kastner \& Bhatia (1992)]{kast92}Kastner, S. O., \& Bhatia, A. K. 1992, ApJ, 398, 698
\bibitem[Kraemer et al.(1998a)]{krae98a}Kraemer, S. B., Ruiz, J. R., \& Crenshaw, D. M. 1998, ApJ, 508, 232
\bibitem[Kraemer et al.(1998b)]{krae98b}Kraemer, S. B., et al., 1998, ApJ, 499, 719
\bibitem[Kraemer et al.(2001)]{krae01}Kraemer, S. B., Crenshaw, D. M., Gabel J. R. 2001a, \apj, 557, 30 (KC01)
\bibitem[Kraemer et al.(2002)]{krae02}Kraemer, S. B., Crenshaw, D. M., George, I. M., Netzer, H., Turner, T. J., \& Gabel J. R. 2002, \apj, 577, 98
\bibitem[Kraemer et al.(2003)]{krae03}Kraemer, S. B., et al. 2003, \apj, 582, 125
\bibitem[Kriss (2002)]{kris02}Kriss, G. A.  2002, in ASP Conf. Ser. 255, Mass Outflow in Active Galactic Nuclei: New Perspectives, ed. D. M. Crenshaw, S. B. Kraemer, \& I. M. George (San Francisco: ASP), 69
\bibitem[Krolik \& Kriss (1995)]{krol95}Krolik, J. H., \& Kriss, G. A. 1995, \apj, 447, 512
\bibitem[Krolik \& Kriss (2001)]{krol01}Krolik, J. H., \& Kriss, G. A. 2001, \apj, 561, 684
\bibitem[Krongold et al.(2005)]{kron05}Krongold, Y., Nicastro, F., Brickhouse, N. S., Elvis, M., \& S. Mathur 2005, ApJ, 622, 842
\bibitem[Lindler (1999)]{lind99}Lindler, D. 1999 CALSTIS Reference Guide (CALSTIS Version 5.1)
\bibitem[Mathur et al.(1994)]{math94}Mathur, S., Wilkes, B., Elvis, M., \& Fiore, F. 1994, ApJ, 434, 493
\bibitem[Mathur et al.(1995)]{math95}Mathur, S., Elvis, M., \& Wilkes, B. 1995, ApJ, 452, 230
\bibitem[Morton (1991)]{mort91}Morton, D. C. 1991, \apjs, 77, 119
\bibitem[Netzer et al.(2003)]{netz03}Netzer, H., et al. 2003, ApJ, 599, 933 (Paper IV)
\bibitem[Onken \& Peterson (2002)]{onke02}Onken, C. A., \& Peterson, B. M.  2002, \apj, 572, 746
\bibitem[Peterson \& Wandel (1999)]{pete99}Peterson, B. M., \& Wandel, A., 1999, ApJ, 521, L95
\bibitem[Peterson et al.(2004)]{pete04}Peterson, B. M., et al., 2004, ApJ, 613, 682
\bibitem[Rees (1987)]{rees87}Rees, M. J. 1987, MNRAS, 228, 47
\bibitem[Reichert et al.(1994)]{reic94}Reichert, G. A., et al.\  1994, \apj, 425, 582
\bibitem[Reynolds (1997)]{reyn97}Reynolds, C. S. 1997, MNRAS, 286, 513
\bibitem[Savage \& Sembach (1991)]{sava91}Savage, B. D., \& Sembach, K. R. 1991, \apj, 379, 245
\bibitem[Wampler et al.(1993)]{wamp93}Wampler, E. J., Bergeron, J., \& Petitjean, P. 1993, A\&A, 273, 15
\bibitem[Ward \& Morris (1984)]{ward84}Ward, M. \& Morris, S. 1984, MNRAS, 207, 867
\bibitem[Wills et al.(1993)]{will93}Wills, B., et al. 1993, ApJ, 410, 534
\end{thebibliography}
\end{document}